\documentclass{LMCS}

\usepackage{enumerate}
\usepackage{bm}
\usepackage{amsmath,amssymb}
\usepackage{mathpartir}
\usepackage{stmaryrd}

\newcommand{\al}{\ensuremath{\alpha}}

\newcommand{\aML}{\al{}ML}

\newcommand{\Chap}[1]{Chapter~{#1}}

\newcommand{\Conv}[1]{Convention~{#1}}
\newcommand{\Cor}[1]{Corollary~{#1}}
\newcommand{\Defn}[1]{Definition~{#1}}

\newcommand{\Fig}[1]{Figure~{#1}}
\newcommand{\Lem}[1]{Lemma~{#1}}

\newcommand{\Prop}[1]{Proposition~{#1}}
\newcommand{\Rem}[1]{Remark~{#1}}
\newcommand{\Sect}[1]{Section~{#1}}

\newcommand{\Thm}[1]{Theorem~{#1}}

\newcommand{\aProlog}{\ensuremath{\alpha}Prolog}

\newcommand{\act}{\cdot}

\newcommand{\bigsum}[2]{\sum_{#1}{#2}}

\newcommand{\bimp}{\Longleftrightarrow}
\newcommand{\bnf}{::=}

\newcommand{\cartesian}[2]{{#1}\times{#2}}

\newcommand{\comp}{\circ}
\renewcommand{\conj}{\wedge} 
\newcommand{\CONJRANGE}[2]{{#1}\AND\cdots\AND{#2}}

\newcommand{\defeq}{\triangleq}
\newcommand{\den}[1]{\llbracket{#1}\rrbracket}

\newcommand{\disu}{\uplus}
\newcommand{\disurange}[2]{{#1}\disu\cdots\disu{#2}}
\newcommand{\dom}[1]{\mathit{dom}\pn{#1}}

\newcommand{\ent}{\vdash}

\newcommand{\equantb}[2]{\exists{#1}\pn{#2}}

\newcommand{\frenv}{\nabla}
\newcommand{\freshfor}[2]{{#1}\metahash{#2}}

\newcommand{\imp}{\Longrightarrow}

\newcommand{\inter}{\cap}
\newcommand{\kw}[1]{\text{\normalfont\texttt{#1}}}

\newcommand{\lrange}[2]{{#1},\ldots,{#2}}

\newcommand{\metahash}{
  \ensuremath{\mathrel{%
      \hbox{\ooalign{%
          \hfil$\approx$\hfil\cr%
          \hfil$\sslash$\hfil}}}}}

\newcommand{\Nat}{\mathbb{N}}

\newcommand{\pair}[2]{\pn{#1\mathbin{,}#2}}
\newcommand{\pn}[1]{({#1})}

\newcommand{\qtn}[1]{``{#1}''}
\newcommand{\qtup}[4]{\pn{{#1},{#2},{#3},{#4}}}
\newcommand{\quant}[3]{#1 #2.\ #3}

\newcommand{\set}[1]{\overline{#1}}
\newcommand{\setb}[1]{\{{#1}\}}
\newcommand{\setrange}[2]{\setb{\lrange{#1}{#2}}}
\newcommand{\setwhere}[2]{\setb{{#1}\mid{#2}}}
\newcommand{\msetb}[1]{\{\!\{{#1}\}\!\}}
\newcommand{\msetwhere}[2]{\msetb{{#1}\mid{#2}}}

\newcommand{\sset}{\mathcal{S}}

\newcommand{\twinkle}[1]{#1^{*}}

\newcommand{\uquant}[2]{\quant{\forall}{#1}{#2}}

\newcommand{\triple}[3]{\pn{{#1},{#2},{#3}}}

\newcommand{\constr}{c}
\newcommand{\Constr}[1]{\mathit{Constr}_{#1}}

\renewcommand{\prob}{\set{\constr}}
\newcommand{\Prob}[1]{\mathit{Prob}_{#1}}
\newcommand{\SIG}{\Sigma}
\newcommand{\SIGCons}[1]{\mathbb{C}_{#1}}
\newcommand{\SIGDtys}[1]{\mathbb{D}_{#1}}
\newcommand{\SIGNtys}[1]{\mathbb{N}_{#1}}
\newcommand{\term}{t}
\newcommand{\Term}[1]{\mathit{Term}_{#1}}

\newcommand{\var}{x}
\newcommand{\Var}{\mathit{Var}}

\newcommand{\vars}[1]{\mathit{vars}\pn{#1}}
\newcommand{\varset}{\set{\var}}
\newcommand{\vare}{\set{\var}}

\newcommand{\vu}{u}
\newcommand{\vv}{v}
\newcommand{\vw}{w}
\newcommand{\vx}{x}
\newcommand{\vy}{y}
\newcommand{\vz}{z}
\newcommand{\ABS}[2]{\kw{<}#1\kw{>}#2}

\newcommand{\DISTINCT}[1]{{\large\FRESHFOR}_{#1}}

\newcommand{\AND}{\mathrel{\kw{\&}}}
\newcommand{\APP}[2]{{#1}\,{#2}}

\newcommand{\ARROW}{\rightarrow}

\newcommand{\COMMA}{\kw{,}}
\newcommand{\CON}{K}
\newcommand{\CONJ}[2]{{#1}\AND{#2}}

\newcommand{\ECONSYM}{\EQ}
\newcommand{\ECON}[2]{{#1}\ECONSYM{#2}}
\newcommand{\EQ}{\mathbin{\text{\texttt{=}}}}

\newcommand{\FCONSYM}{\FRESHFOR}
\newcommand{\FCON}[2]{{#1}\FCONSYM{#2}}

\newcommand{\FRESHFOR}{\mathbin{\kw{\#}}}

\newcommand{\OFTY}{\kw{:}}

\newcommand{\PAIR}[2]
{\mathopen{\kw{(}}{#1}\kw{,}\,{#2}\mathclose{\kw{)}}}

\newcommand{\tuple}[2]{\mathopen{(}#1,\ldots,#2\mathclose{)}}
\newcommand{\TUPLE}[2]
{\mathopen{\kw{(}}#1\COMMA\ldots\COMMA#2\mathclose{\kw{)}}}

\newcommand{\UNIT}{\kw{()}}

\newcommand{\aec}[1]{[#1]_\alpha}
\newcommand{\aeq}[3]{{#1}\aeqsym{#2}\OFTY{#3}}
\newcommand{\aeqnt}[2]{{#1}\aeqsym{#2}}
\newcommand{\aeqsym}{=_{\alpha}}

\newcommand{\Atree}[2]{\ensuremath{\textit{$\alpha$-Tree}_{#1}\pn{#2}}}
\newcommand{\altree}{\mathfrak{a}}
\newcommand{\cprob}[2]{\exists{#1}\pn{#2}}

\newcommand{\fn}[1]{\mathit{FN}\pn{#1}}

\newcommand{\idp}{\iota}

\newcommand{\name}{n}

\newcommand{\Name}{\mathit{Name}}
\newcommand{\names}{\set{\name}}
\newcommand{\namesof}[1]{\mathit{names}\pn{#1}}
\newcommand{\NonPermSat}{\textsf{NonPermSat}}

\newcommand{\notrans}{\not\trans}
\newcommand{\NP}{\textsf{NP}}
\newcommand{\pact}[2]{{#1}\act{#2}}
\newcommand{\pcomp}[2]{{#1}\comp{#2}}
\newcommand{\pcomps}[3]{\pcomp{#1}{\pcomp{#2}{#3}}}
\newcommand{\pcompss}[4]{\pcomp{#1}{\pcomps{#2}{#3}{#4}}}

\newcommand{\perm}{\pi}
\newcommand{\Perm}{\mathit{Perm}}

\newcommand{\pinv}[1]{{#1}^{-1}}

\newcommand{\REL}{R}
\newcommand{\sat}{\models}
\newcommand{\SAT}[1]{{}\sat{#1}}
\newcommand{\SATIS}[2]{{#1}\sat{#2}}

\newcommand{\size}[1]{\sizefn\pn{#1}}
\newcommand{\sizefn}{\mathit{size}}
\newcommand{\sort}[1]{\sortfn\pn{#1}}
\newcommand{\sortfn}{\mathit{sort}}
\renewcommand{\succ}[1]{\succword\pn{#1}}
\newcommand{\succword}{\mathit{succ}}
\newcommand{\swap}[2]{\pn{{#1}\,{#2}}}

\newcommand{\trans}{\longrightarrow}
\newcommand{\tree}{g}
\newcommand{\Tree}[2]{\mathit{Tree}_{#1}\pn{#2}}
\newcommand{\TreeNoType}[1]{\mathit{Tree}_{#1}}
\newcommand{\vapp}[2]{\den{#2}_{#1}}
\newcommand{\vln}{V}

\newcommand{\vlnw}{W}
\newcommand{\ABSTY}[2]{\kw{[}#1\kw{]}#2}

\newcommand{\CONTYPE}[4]{\pn{#1\OFTY\FUNTY{#2}{#3}}\in#4}
\newcommand{\CROSS}{\kw{*}}

\newcommand{\dty}{D}
\newcommand{\etenv}{\Delta}

\newcommand{\ok}{\mathit{\ ok}}
\newcommand{\FUNTY}[2]{{#1}\ARROW{#2}}

\newcommand{\nty}{N}

\newcommand{\PRODTY}[2]{#1\mathbin{\CROSS}#2}

\newcommand{\TUPTY}[2]
{#1\mathbin{\CROSS}\cdots\mathbin{\CROSS}#2}
\newcommand{\ty}{T}
\newcommand{\Ty}[1]{\mathit{Ty}_{#1}}
\newcommand{\UNITTY}{\kw{unit}}

\newcommand{\sapp}[2]{{#2}[#1]}
\newcommand{\sub}[2]{{#2}/{#1}}
\newcommand{\subst}{\sigma}

\newcommand{\lamterm}{M}

\newcommand{\SKW}{\kw{S}}

\newcommand{\SUSP}[2]{{#1}\,{#2}}

\newcommand{\uvar}{X}

\newcommand{\transrule}[2]{\pn{{#1}{#2}}}
\newcommand{\erule}[1]{\transrule{E}{#1}}
\newcommand{\fcons}[4]{\FCON{#1}{\varss{#2}{#3}{#4}}}
\newcommand{\frule}[1]{\transrule{F}{#1}}
\newcommand{\varss}[3]{{#1}_{{#2}\cdot\cdot{#3}}}
\newcommand{\abs}[2]{\ABS{#1}{#2}}
\newcommand{\abss}[4]{\abs{\varss{#1}{#2}{#3}}{#4}}

\newcommand{\narrow}[4]{[{#1};{#2}]\narrowsym[{#3};{#4}]}
\newcommand{\narrowsym}{\Longmapsto}
\newcommand{\nrule}[1]{\transrule{N}{#1}}

\newcommand{\Natdty}{\kw{nat}}
\newcommand{\fl}[1]{\ensuremath{{#1}^{\flat}}}
\newcommand{\flb}[2]{\ensuremath{{#2}_{#1}^{\flat}}}
\newcommand{\expjb}[4]{{#2}\ent_{#1}{#3}\OFTY{#4}}
\newcommand{\formjb}[3]{{#2}\ent_{#1}{#3}\mathit{\ ok}}
\newcommand{\NOTSAT}[1]{{}\not\sat{#1}}
\newcommand{\UNITS}[2]{\PAIR{\UNIT^{#1}}{#2}}
\newcommand{\NatPlus}{\Nat_{+}}
\newcommand{\termeq}[1]{\lceil{#1}\rceil}
\newcommand{\termsize}[2]{\termeq{#1}\pn{#2}}

\newcommand{\NatMS}{\mathcal{M}}
\newcommand{\meassym}{\mu}
\newcommand{\meas}[2]{\meassym\pn{#1}\pn{#2}}
\newcommand{\measa}[2]{\meassym_1\pn{#1}\pn{#2}}
\newcommand{\measb}[2]{\meassym_2\pn{#1}\pn{#2}}

\newcommand{\mordsym}{\prec}

\newcommand{\mordsubsym}[1]{\mordsym_{#1}}
\newcommand{\mordsub}[3]{{#2}\mordsubsym{#1}{#3}}

\newcommand{\nvars}[1]{\mathit{nvars}\pn{#1}}

\newcommand{\Varb}[1]{\Var_{\textit{#1}}}
\newcommand{\VarName}{\Varb{Name}}
\newcommand{\VarTemp}{\Varb{Temp}}

\newcommand{\VarNvar}{\Varb{Nvar}}
\newcommand{\VarPvar}[1]{\Var_{#1}}
\newcommand{\Nvar}{\mathit{Nvar}}
\newcommand{\va}{a}
\newcommand{\vb}{b}
\newcommand{\vara}{A}
\newcommand{\varae}{\set{\vara}}
\newcommand{\varas}[1]{\mathit{namevars}\pn{#1}}
\newcommand{\varb}{B}
\newcommand{\pvar}{Q}
\newcommand{\pvare}{\set{\pvar}}
\newcommand{\pvars}[1]{\mathit{pvars}\pn{#1}}
\newcommand{\PERM}{\Pi}
\newcommand{\euvln}{\theta}
\newcommand{\euesym}{\approx}
\newcommand{\euecon}[2]{{#1}\euesym{#2}}
\newcommand{\eufsym}{\mathrel{\#}}
\newcommand{\eufcon}[2]{{#1}\eufsym{#2}}
\newcommand{\euprob}{S}
\newcommand{\eusem}[1]{\mathit{Sat}\pn{#1}}
\newcommand{\consistency}[3]{\mathbb{C}\triple{#1}{#2}{#3}}
\newcommand{\bijcon}[4]{\ECON{\ABS{#1}{\ABS{#2}{#1}}}{\ABS{#3}{\ABS{#4}{#3}}}}
\newcommand{\swapcon}[4]{\ECON{\ABS{#1}{\ABS{#2}{#3}}}{\ABS{#2}{\ABS{#1}{#4}}}}
\newcommand{\vvar}[1]{\var_{#1}}
\newcommand{\pvvar}[2]{\var_{\pair{#1}{#2}}}
\newcommand{\eutrn}[1]{\mathtt{tr}\pn{#1}}
\newcommand{\eutrnterm}[5]{\eutrn{#1}_{#2}=\equantb{#5}{#3\textsf{\ where\ }#4}}
\newcommand{\eutrnconstr}[4]{\eutrn{#1}_{#2}=\equantb{#4}{#3}}
\newcommand{\eutrnprob}[3]{\eutrn{#1}=\equantb{#3}{#2}}
\newcommand{\euvlntrn}[2]{\vln_{\pair{#1}{#2}}}

\newcommand{\eujmt}[4]{{#1};{#2};{#3}\ent{#4}\ok}

\newcommand{\numof}[1]{k_{#1}}
\newcommand{\eunenv}[4]{\etenv_{\qtup{#1}{#2}{#3}{#4}}}

\def\doi{7 (3:06) 2011}
\lmcsheading%
{\doi}
{1--31}
{}
{}
{Jan.~\phantom03, 2011}
{Aug.~25, 2011}
{}

\begin{document}

\title[Non-permutative nominal constraints]{Constraint solving in non-permutative nominal abstract syntax}

\author[M.~R.~Lakin]{Matthew R. Lakin}	
\address{University of Cambridge Computer Laboratory, Cambridge CB3
  0FD, UK, and Department of Computer
Science, University of New Mexico, Albuquerque, NM 87131}	
\email{Matthew.Lakin@cl.cam.ac.uk}  
\thanks{\emph{Research supported by:} UK EPSRC grant EP/D000459/1.}	




\begin{abstract}
  \noindent Nominal abstract syntax is a popular first-order technique for
encoding, and reasoning about, abstract syntax involving binders.
Many of its applications involve \emph{constraint solving}.
The most commonly used constraint solving algorithm over nominal
abstract syntax is the Urban-Pitts-Gabbay \emph{nominal unification}
algorithm, which is well-behaved, has a well-developed theory and is
applicable in many cases.
However, certain problems require a constraint solver
which respects the equivariance property of nominal logic, such as Cheney's
\emph{equivariant unification} algorithm.
This is more powerful but is more complicated and computationally hard.
In this paper we present a novel algorithm for solving constraints
over a simple variant of nominal abstract syntax which we call
\emph{non-permutative}.
This constraint problem has similar complexity to
equivariant unification but without many of the additional
complications of the equivariant unification term language.
We prove our algorithm correct, paying particular attention to
issues of termination, and present an explicit translation
of name-name equivariant unification problems into
non-permutative constraints.
\end{abstract}

\maketitle

\section{Introduction}
\label{sec:intro}

Constraint solving over the abstract syntax of programming languages
is vital in many areas of logic and computer science.
For example, many compiler optimisations are typically phrased as constraint problems.
The abstract syntax in question often involves binding
constructs, such as $\lambda$-expressions or $\forall$-quantifiers.
In these cases we would want the abstract syntax encoding to
respect \al-equivalence of binding structures---this is known
as the Barendregt variable convention \cite{BarendregtHP:lamcss}.

One approach to representing and manipulating abstract syntax with
binders is \emph{nominal abstract syntax}
\cite{PittsAM:newaas-jv}, which was developed as a first-order
theory of abstract syntax involving bound names.
The theory is based on permuting, rather than substituting, names,
as this has more convenient logical properties.
Bindable names in the object-language are represented by
meta-level names ranged over by $\name$
\pn{these are often called \emph{atoms} in the literature}.
These names follow Gabbay's \emph{permutative convention}
\cite{GabbayMJ:capasn-jv} which states that distinct meta-variables
$\name_1$ and $\name_2$ over names always denote distinct names.
The object-level binding of a name $\name$ in a term $\term$ is
is represented by the abstraction term $\ABS{\name}{\term}$.
This is not itself a binder, so $\ABS{\name_1}{\name_1}$ and
$\ABS{\name_2}{\name_2}$ are considered to be distinct terms
if $\name_1\neq\name_2$ \pn{though they should behave similarly
as they are \al-equivalent}.

Logic programming over nominal abstract syntax requires
a unification algorithm which can unify nominal terms
modulo \al-equivalence.
One such algorithm is \emph{nominal unification} \cite{PittsAM:nomu-jv},
which is a simple, well-studied constraint solving algorithm
for nominal abstract syntax.
The algorithm extends first-order unification to work on nominal terms
modulo \al-equivalence by adding \emph{freshness} constraints
which are satisfied if a given name does not appear free in a term.
This allows terms involving binders to be equated by checking
that they have the same topology of name bindings.
Nominal unification enjoys unique, most general solutions
\cite[\Thm{3.7}]{PittsAM:nomu-jv} and is known to be
decidable in quadratic time \cite{LevyJ:nomuho,LevyJ:effnua,CalvesC:fironl}.

Resolution using nominal unification is the basis of the
\aProlog{} nominal logic programming language \cite{CheneyJ:alpplp}.
However, here we encounter a problem---resolution using
nominal unification is incomplete for nominal logic
\cite{PittsAM:nomlfo-jv}.
The issue is that nominal unification does not respect the
\emph{equivariance} property of nominal logic, that is,
closure under name-permutations.
A standard example, taken from \cite{CheneyJ:equu-jv},
concerns capture-avoiding substitution over $\lambda$-terms
encoded in nominal abstract syntax.
Writing $\mathtt{subst}\triple{\lamterm}{\lamterm'}{\name}$
to represent the capture-avoiding substitution function
$\sapp{\sub{\name}{\lamterm'}}{\lamterm}$,
the following two rewrite rules implement the case
when $\lamterm$ is a variable.
\begin{mathpar}
  \mathtt{subst}\triple{\mathtt{var}\pn{\name}}{\lamterm'}{\name}
  \rightarrow\lamterm' \and
  \mathtt{subst}\triple{\mathtt{var}\pn{\name'}}{\lamterm'}{\name}
  \rightarrow\mathtt{var}\pn{\name'}
\end{mathpar}
From the first rule we infer that $\mathtt{subst}
\triple{\mathtt{var}\pn{\name}}{\mathtt{var}\pn{\name'}}{\name}
\rightarrow\mathtt{var}\pn{\name'}$.
However, nominal unification cannot compute a substitution for
$\lamterm'$ such that
\begin{equation*}
  \aeqnt
  {\mathtt{subst}\triple{\mathtt{var}\pn{\name}}{\lamterm'}{\name}}
  {\mathtt{subst}\triple{\mathtt{var}\pn{\name'}}
  {\mathtt{var}\pn{\name}}{\name'}}.
\end{equation*}
In nominal logic this equation holds modulo a permutation,
and hence nominal unification does not suffice for complete
proof search in all cases.

A workaround is to define a well-formedness condition on
\aProlog{} programs to isolate those which can be executed
correctly using nominal unification \cite{UrbanC:avoeap}.
A more general solution is to use a more powerful constraint
solving algorithm which takes equivariance into account,
such as Cheney's \emph{equivariant unification} algorithm
\cite{CheneyJ:equu-jv}.
Equivariant unification generalises the term language of
nominal unification to include name variables $\vara$
which stand for unknown names
\pn{and which do not follow the permutative convention}
and permutation variables $\pvar$ which stand for
unknown permutations.
The syntax of names and permutations in equivariant unification
is non-trivial---for example, the terms involved in a swapping
may themselves contain nested swappings!
Furthermore, these compound name expressions may appear in abstraction
position, so we can write terms such as
\begin{equation*}
  \ABS{\SUSP{\swap{\pn{\SUSP{\pn{\pcomp{\pvar}{\pvar'}}}
  {\vara}}}{\pn{\SUSP{\pinv{\pvar}}{\vara}}}}{\name}}
  {\pn{\SUSP{\pvar'}{\name}}}
\end{equation*}
even though the meaning of such a term is by no means obvious.
The equivariant unification algorithm uses \qtn{permutation graphs}
to solve generalised equality and freshness constraints.
This constraint problem is \NP{}-hard \cite{CheneyJ:equu-jv},
and there are no longer unique, most general solutions.
However, the main issue with equivariant unification are that
the term language is complicated \pn{see below},
making it difficult to implement the algorithm
and interpret the resulting answers.

\subsection{Contributions}
\label{subsec:contributions}

In this paper we present an alternative to equivariant unification,
in the form of a constraint algorithm over \emph{non-permutative
nominal abstract syntax} \pn{NPNAS}.
This is a mild generalisation of standard nominal abstract syntax
and is a syntactic subset of the equivariant unification term language.
We use the term \emph{non-permutative} because no meta-variables,
not even those representing names in binding position,
follow the permutative convention.
The contributions of this paper are as follows:
\begin{iteMize}{$\bullet$}
  \item the presentation of non-permutative nominal abstract syntax
    as a simple yet powerful extension of existing nominal techniques;
  \item a novel decision procedure for solving equality and freshness
    constraints over non-permutative nominal terms,
    and a proof of its correctness; and
  \item a reduction of name-name equivariant
    unification problems to non-permutative nominal constraints.
\end{iteMize}
We do not address the fact that equivariant unification
is \NP{}-hard, since the NPNAS constraint problem is also
\NP{}-hard.
However, the simplicity of the NPNAS term language and algorithm
offer practical advantages over equivariant unification
when it comes to implementing a nominal logical
programming or rewriting system.
The constraint solving algorithm described in this paper can
be used to implement sound and complete resolution over
inductive definitions involving binders,
as described in \cite{LakinMR:exemli}.
This is an important result given the ubiquity of binders
in logic and computer science.

The rest of this paper is organised as follows.
\Sect{\ref{sec:background}} presents important background
on nominal abstract syntax, nominal unification and
equivariant unification.
In \Sect{\ref{sec:syntax-semantics}} we present the syntax
of NPNAS terms and constraints and define their semantics.
We present a constraint transformation algorithm in
\Sect{\ref{sec:trans}} and use it to derive a correct
decision procedure in \Sect{\ref{sec:correctness}},
paying particular attention to termination.
We address the relationship between NPNAS and equivariant
unification in \Sect{\ref{sec:equivariant}} by defining
an explicit translation of name-name equivariant
unification problems into NPNAS.
We discuss related and future work in \Sect{\ref{sec:related-future}}
and conclude in \Sect{\ref{sec:conc}}.

\section{Background}
\label{sec:background}

This section presents the basics of nominal abstract syntax,
nominal terms, nominal unification and equivariant unification.
This suffices to demonstrate the relationship of the non-permutative
terms and constraints studied in this paper to existing work.
We refer the reader to \cite{PittsAM:nomlfo-jv} for a full
introduction to nominal logic, to \cite{PittsAM:nomu-jv} for
details on nominal unification and \cite{CheneyJ:equu-jv} for
an in-depth treatment of equivariant unification.

\subsection{Nominal abstract syntax}
\label{subsec:background-nas}

As is standard in the world of nominal techniques, we use
\emph{nominal signatures} \cite{PittsAM:nomu-jv}
to specify binding structures in the object-language.

\begin{defi}[Nominal signatures]
\label{def:nomsig}
  A nominal signature $\SIG$ consists of:
  \begin{iteMize}{$\bullet$}
    \item a finite set $\SIGNtys{\SIG}$ of \emph{name sorts},
          ranged over by $\nty$;
    \item a finite set $\SIGDtys{\SIG}$ of \emph{data sorts},
          disjoint from $\SIGNtys{\SIG}$ and ranged over by $\dty$; and
    \item a finite set $\SIGCons{\SIG}$ of \emph{constructors}
          $\CON\OFTY\FUNTY{\ty}{\dty}$, where the
          argument type $\ty\in\Ty{\SIG}$ is generated by the following grammar.
          \begin{displaymath}
            \begin{array}{rcl@{\qquad}l}
              \ty\in\Ty{\SIG} & \bnf
                     & \dty               & \text{\pn{data sorts}} \\
              & \mid & \nty               & \text{\pn{name sorts}} \\
              & \mid & \ABSTY{\nty}{\ty}  & \text{\pn{name abstractions}} \\
              & \mid & \TUPTY{\ty}{\ty}   & \text{\pn{tuples}} \\
              & \mid & \UNITTY            & \text{\pn{unit}}
            \end{array}
          \end{displaymath}
  \end{iteMize}\smallskip
\end{defi}

\noindent For simplicity we will assume that $\SIG$ is such that every type
$\ty\in\Ty{\SIG}$ is \emph{inhabited} by some ground tree
\pn{defined below}.
This property of nominal signatures can be checked
straightforwardly---see \cite[\Sect{3.3.2}]{LakinMR:exemli}
for details.

In standard approaches to nominal abstract syntax
\cite{PittsAM:newaas-jv}, bindable names are represented
explicitly in the syntax of ground syntax trees.
We fix a countably infinite set $\Name$ of names
to stand for object-language names which may be bound.
The meta-variable $\name$ ranges \emph{permutatively} over these.
We assume the existence of a total function $\sortfn$ which
maps every name $\name$ to a name sort $\nty\in\SIGNtys{\SIG}$
such that there are infinitely many names assigned to every name sort.
We say that $\name\in\Name\pn{\nty}$ if $\sort{\name}=\nty$.

\begin{defi}[Ground trees]
\label{def:trees}
  We write $\TreeNoType{\SIG}$ for the set of all syntax trees
  over the nominal signature $\SIG$.
  We refer to these as \emph{ground trees} following the
  terminology of \cite{LakinMR:exemli}, though they are
  often referred to as \pn{ground} nominal terms.
  With names \pn{and unit} as our building blocks, we define classes 
  $\tree\in\Tree{\SIG}{\ty}$ of syntax trees of the various
  types by constructor application, tupling and name abstraction,
  as follows.
  \begin{mathpar}
    \inferrule{\sort{\name}=\nty}{\name\in\Tree{\SIG}{\nty}} \and
    \inferrule{\ }{\UNIT\in\Tree{\SIG}{\UNITTY}} \and
    \inferrule{\tree_1\in\Tree{\SIG}{\ty_1} \\ \cdots \\
    \tree_k\in\Tree{\SIG}{\ty_k}}
    {\TUPLE{\tree_1}{\tree_k}\in\Tree{\SIG}{\TUPTY{\ty_1}{\ty_k}}} \and
    \inferrule{\tree\in\Tree{\SIG}{\ty} \\
    \CONTYPE{\CON}{\ty}{\dty}{\SIG}}
    {\APP{\CON}{\tree}\in\Tree{\SIG}{\dty}} \and
    \inferrule{\sort{\name}=\nty \\ \tree\in\Tree{\SIG}{\ty}}
    {\ABS{\name}{\tree}\in\Tree{\SIG}{\ABSTY{\nty}{\ty}}}
  \end{mathpar}
\end{defi}

The abstraction $\ABS{\name}{\tree}$ represents
a term with a bound name.
This term-former is not regarded as a binder,
which means that, for distinct names $\name$ and $\name'$,
we regard $\ABS{\name}{\name}$ and $\ABS{\name'}{\name'}$
as distinct ground trees.
This is a consequence of following Gabbay's
\emph{permutative convention} \cite{GabbayMJ:capasn-jv}.

\begin{defi}[Permutations and permutation actions]
\label{def:perms}
  Let $\Perm$ be the set of all finite permutations over
  $\Name$, that is, the set of all bijections $\perm$ such that
  $\perm\pn{\name}=\name$ for all but finitely many $\name$.
  Any element of $\Perm$ can be represented as a finite
  list of name-swappings of the form $\swap{\name}{\name'}$.
  The \emph{action} of a permutation $\perm$ on a ground
  tree $\tree$ is to rename all names appearing in $\tree$
  \pn{including those in abstraction position} according to
  $\perm$. This is defined as follows.
  \begin{mathpar}
    \pact{\perm}{\name}\defeq\perm\pn{\name} \and
    \pact{\perm}{\UNIT}\defeq\UNIT \and
    \pact{\perm}{\TUPLE{\tree_1}{\tree_k}}\defeq
    \TUPLE{\pact{\perm}{\tree_1}}{\pact{\perm}{\tree_k}} \and
    \pact{\perm}{\pn{\APP{\CON}{\tree}}}\defeq
    \APP{\CON}{\pn{\pact{\perm}{\tree}}} \and
    \pact{\perm}{\pn{\ABS{\name}{\tree}}}\defeq
    \ABS{\perm\pn{\name}}{\pn{\pact{\perm}{\tree}}}
  \end{mathpar}
\end{defi}

Using the definition of permutation action,
\Fig{\ref{fig:trees-equality-freshness}} defines
a type-directed equality relation between two ground trees
and a freshness relation between a name and a ground tree.
The equality relation corresponds
\pn{$\aeq{\tree}{\tree'}{\ty}$} to
\al-equivalence \cite{PittsAM:newaas-jv}.
This definition of \al-equivalence paraphrases that of
\cite{BarendregtHP:lamcss}, as shown by Gabbay and Pitts
\cite[\Prop{2.2}]{PittsAM:newaas-jv}.
The freshness relation \pn{$\freshfor{\name}{\tree}$}
holds when the name $\name$ is not free in the
ground tree $\tree$.
We write $\fn{\tree}$ for the set of names $\name$
which appear in $\tree$ and are such that
$\freshfor{\name}{\tree}$ holds.

Finally, a relation $\REL\subseteq\TUPTY{\Tree{\SIG}{\ty_1}}{\Tree{\SIG}{\ty_k}}$
is \emph{equivariant} if, for all permutations $\perm$,
$\REL\tuple{\tree_1}{\tree_k}$ holds iff
$\REL\tuple{\pact{\perm}{\tree_1}}{\pact{\perm}{\tree_k}}$ holds.
It is not hard to show that the equality and freshness
relations from \Fig{\ref{fig:trees-equality-freshness}}
are both equivariant.

\begin{figure}
  \mbox{\begin{mathpar}
    \inferrule{\sort{\name}=\nty}{\aeq{\name}{\name}{\nty}} \and
    \inferrule{\ }{\aeq{\UNIT}{\UNIT}{\UNITTY}} \and
    \inferrule{\aeq{\tree_1}{\tree_1'}{\ty_1} \\ \cdots \\
    \aeq{\tree_k}{\tree_k'}{\ty_k}}
    {\aeq{\TUPLE{\tree_1}{\tree_k}}{\TUPLE{\tree_1'}{\tree_k'}}
    {\TUPTY{\ty_1}{\ty_k}}} \and
    \inferrule{\aeq{\tree}{\tree'}{\ty} \\
    \CONTYPE{\CON}{\ty}{\dty}{\SIG}}
    {\aeq{\APP{\CON}{\tree}}{\APP{\CON}{\tree'}}{\dty}} \and
    \inferrule{\aeq{\tree}{\tree'}{\ty} \\ \sort{\name}=\nty}
    {\aeq{\ABS{\name}{\tree}}{\ABS{\name}{\tree'}}{\ABS{\nty}{\ty}}} \and
    \inferrule{\sort{\name}=\sort{\name'}=\nty \\\\
    \name\neq\name' \\ \freshfor{\name}{\tree'} \\
    \aeq{\tree}{\pact{\swap{\name}{\name'}}{\tree'}}{\ty}}
    {\aeq{\ABS{\name}{\tree}}{\ABS{\name'}{\tree'}}
    {\ABSTY{\nty}{\ty}}} \and
    \inferrule{\name\neq\name'}{\freshfor{\name}{\name'}} \and
    \inferrule{\ }{\freshfor{\name}{\UNIT}} \and
    \inferrule{\freshfor{\name}{\tree_1} \\ \cdots \\
    \freshfor{\name}{\tree_k}}
    {\freshfor{\name}{\TUPLE{\tree_1}{\tree_k}}} \and
    \inferrule{\freshfor{\name}{\tree}}
    {\freshfor{\name}{\APP{\CON}{\tree}}} \and
    \inferrule{\ }{\freshfor{\name}{\ABS{\name}{\tree}}} \and
    \inferrule{\name\neq\name' \\ \freshfor{\name}{\tree}}
    {\freshfor{\name}{\ABS{\name'}{\tree}}}
  \end{mathpar}}
  \caption{Equality and freshness for ground trees}
  \label{fig:trees-equality-freshness}
\end{figure}

\subsection{Nominal unification}
\label{subsec:background-nu}

\emph{Nominal unification} \cite{PittsAM:nomu-jv} is a simple, well-studied
constraint solving algorithm which extends first-order unification
to work on ground trees with binders modulo \al-conversion.
The language of ground trees from \Defn{\ref{def:trees}} is extended
to include metavariables with suspended permutations $\SUSP{\perm}{\uvar}$.
When $\uvar$ is instantiated by a substitution $\subst$ the permutation
must be applied to produce the result $\pact{\perm}{\pn{\subst\pn{\uvar}}}$.
Problems consist of equality \pn{$\ECON{\term}{\term'}$} and
freshness \pn{$\FCON{\name}{\term'}$} constraints.
These are solved in the context of a \emph{freshness environment}
$\frenv$ of freshness assumptions $\FCON{\name}{\uvar}$ between a
name and a meta-variable which constrain the free names in an
unknown term: if $\pn{\FCON{\name}{\uvar}}\in\frenv$ then $\uvar$
cannot be replaced by any term which has a free occurrence of $\name$.

\subsection{Equivariant unification}
\label{subsec:background-eu}

As mentioned above, \emph{equivariant unification}
\cite{CheneyJ:equu-jv} considerably extends the
term language of nominal unification,
with complex permutation expressions involving
unknown names and unknown permutations.
For the purposes of this paper it suffices to
consider equivariant unification problems involving only
terms of a fixed name sort $\nty$, ranged over by $\va$.
These are generated by the grammar below.
We write $\name$ where \cite{CheneyJ:equu-jv} uses $\mathsf{a}$,
for consistency with the rest of this paper.
\begin{displaymath}
  \begin{array}{r@{\quad}rcl@{\qquad}l}
    \text{Vertices} & \vv,\vw & \bnf
            & \name & \text{\pn{name}} \\
    && \mid & \vara & \text{\pn{name variable}} \\[\jot]
    \text{Name-terms} & \va,\vb & \bnf
            & \pact{\PERM}{\vv} & \text{\pn{suspended permutation}} \\[\jot]
    \text{Permutation-terms} & \PERM & \bnf
            & \idp & \text{\pn{identity}} \\
    && \mid & \swap{\va}{\vb} & \text{\pn{swap}} \\
    && \mid & \pvar & \text{\pn{permutation variable}}
  \end{array}
\end{displaymath}
\qtn{Vertices} is a term used in \cite{CheneyJ:equu-jv},
where equivariant unification problems are represented
as \qtn{permutation graphs}.
A compound permutation expression $\PERM$ may be an unknown
permutation $\pvar$, a swapping $\swap{\va}{\va'}$ or an
explicit permutation composition or inversion.
Equivariant unification name-terms may be either
concrete names $\name$ or name-variables $\vara$.
We abbreviate $\pact{\idp}{\vv}$ as just $\vv$ in most cases.
We write $\eujmt{\names}{\varae}{\pvare}{\va}$ to mean that
$\names\supseteq\namesof{\va}$, $\varae\supseteq\varas{\va}$ and
$\pvare\supseteq\pvars{\va}$, and extend this definition to
other elements of equivariant unification syntax in the obvious way.

The semantics of equivariant unification problems
was defined in \cite{CheneyJ:equu-jv} and we briefly
summarise the relevant details here.
We concern ourselves with ground valuations $\euvln$
applied to name-terms, in particular, the portion of
the valuation that provides values for name variables
$\vara$ and permutation variables $\pvar$,
in terms of ground names $\name$.
\begin{iteMize}{$\bullet$}
  \item If $\vara\in\dom{\euvln}$ then $\euvln\pn{\vara}=\name$, for some
    $\name\in\Name$.
  \item If $\pvar\in\dom{\euvln}$ then $\euvln\pn{\pvar}$ is a
    \emph{ground permutation}, which can be represented as a finite
    \pn{possibly empty} list of name-swappings $\swap{\name}{\name'}$.
\end{iteMize}
The semantics of name-name equivariant unification problems
is as follows, after \cite{CheneyJ:equu-jv}.
\begin{iteMize}{$\bullet$}
  \item $\SATIS{\euvln}{\euecon{\va}{\vb}}$
    iff $\aeq{\euvln\pn{\va}}{\euvln\pn{\vb}}{\nty}$, using the
    rules for \al-equivalence from \Fig{\ref{fig:trees-equality-freshness}}.
  \item $\SATIS{\euvln}{\eufcon{\va}{\vb}}$
    iff $\freshfor{\euvln\pn{\va}}{\euvln\pn{\vb}}$, using the
    rules for freshness from \Fig{\ref{fig:trees-equality-freshness}}.
  \item If $\euprob$ is a finite set of equivariant unification constraints
    $\constr$ \pn{referred to as a \emph{problem}} then
    $\eusem{\euprob}=\setwhere{\euvln}
    {\uquant{\constr\in\euprob}{\SATIS{\euvln}{\constr}}}$.\smallskip
\end{iteMize}

\noindent Name-name equivariant unification problems are known to be
\NP{}-complete, whereas full equivariant unification
\pn{at an arbitrary type $\ty$} is known to be \NP{}-hard,
but not necessarily \NP{}-complete  \cite{CheneyJ:equu-jv}.

The additional constructs supported by equivariant unification
give it the power to solve equations modulo a permutation.
One can compute whether there exists a permutation $\perm$ such that
$\aeqnt{\pn{\pact{\perm}{\term}}}{\term'}$ holds by choosing a fresh
permutation variable $\pvar$ and solving the equivariant unification
problem $\setb{\euecon{\pn{\pact{\pvar}{\term}}}{\term'}}$.
This suffices to allow complete matching and proof-search in
nominal logic programming.

\section{Syntax and semantics of non-permutative constraints}
\label{sec:syntax-semantics}

In this section we present the syntax of non-permutative
nominal terms and constraints over these.
We also define a semantics for non-permutative
nominal constraints.

Schematic terms are used in informal mathematics as templates which may
be used to produce a \pn{potentially infinite} set of ground instances,
quotiented by \al-equivalence.
To permit this, they contain variables which are instantiated with
\pn{\al-equivalence classes of} ground terms according to certain rules.
We fix a countably infinite set $\Var$ of variables as
placeholders for unknown \al-equivalence classes.
We will use various symbols, typically $\var$, $\vy$, etc.,
to range \emph{non-permutatively} over these.

\begin{defi}[Non-permutative nominal terms and atomic constraints]
\label{def:term-constr}
  The sets $\Term{\SIG}$ of \pn{schematic} non-permutative nominal terms
  $\term$ and $\Constr{\SIG}$ of atomic constraints over the
  nominal signature $\SIG$ are defined as follows.
  \begin{displaymath}
    \begin{array}{rcl@{\qquad}l}
      \term\in\Term{\SIG}              & \bnf
             & \var                    & \text{\pn{variable}} \\
      & \mid & \ABS{\var}{\term}       & \text{\pn{abstraction}} \\
      & \mid & \APP{\CON}{\term}       & \text{\pn{data}} \\
      & \mid & \TUPLE{\term}{\term}    & \text{\pn{tuple}} \\
      & \mid & \UNIT                   & \text{\pn{unit}}\\[\jot]
      \constr\in\Constr{\SIG}          & \bnf
             & \ECON{\term}{\term}     & \text{\pn{equality constraint}} \\
      & \mid & \FCON{\var}{\term}      & \text{\pn{freshness constraint}}
    \end{array}
  \end{displaymath}\smallskip
\end{defi}

\noindent Our main departure from traditional approaches to nominal abstract
syntax is that there are no permutative names in the syntax---all
object-level names are represented by non-permutative variables
at the meta-level, even those which appear in binding position.
Since the variables are non-permutative, distinct variables may be
instantiated with the same ground tree, which we call \emph{aliasing}.
As we shall see, the fact that bound names may be aliased means that
a schematic term can be instantiated to multiple
different \al-equivalence classes in general.
The abstraction term-former is \emph{not} a binder,
so there are no meta-level binding constructs in schematic terms.
Hence it is meaningless to define \al-equivalence on schematic
terms directly.

\begin{rem}[Omission of name-constants]
\label{rem:name-constants}
  Unlike the equivariant unification term language,
  we have not included explicit permutative name-constants
  in the grammar of non-permutative nominal terms from
  \Defn{\ref{def:term-constr}}.
  We omit them in part because they do not add expressive power---a
  combination of non-permutative variables and name inequality
  \pn{freshness} constraints can be used to imitate permutative
  name-constants in constraints, as mentioned in
  \Rem{\ref{rem:tractable}}.
  Furthermore, permutative name-constants are not needed to
  achieve a sound and complete encoding of inductive definitions
  over terms involving binders, as shown in
  \cite{LakinMR:residb,LakinMR:exemli}.
\end{rem}

\begin{figure}
  \mbox{\begin{mathpar}
    \inferrule{\var\in\dom{\etenv} \\ \etenv\pn{\var}=\ty}
    {\expjb{\SIG}{\etenv}{\var}{\ty}} \and
    \inferrule{\expjb{\SIG}{\etenv}{\term}{\ty} \\ \CONTYPE{\CON}{\ty}{\dty}{\SIG}}
    {\expjb{\SIG}{\etenv}{\APP{\CON}{\term}}{\dty}} \and
    \inferrule{\ }{\expjb{\SIG}{\etenv}{\UNIT}{\UNITTY}} \and
    \inferrule{\expjb{\SIG}{\etenv}{\term_1}{\ty_1} \\ \cdots \\ \expjb{\SIG}{\etenv}{\term_k}{\ty_k}}
    {\expjb{\SIG}{\etenv}{\TUPLE{\term_1}{\term_k}}{\TUPTY{\ty_1}{\ty_k}}} \and
    \inferrule{\expjb{\SIG}{\etenv}{\var}{\nty} \\ \expjb{\SIG}{\etenv}{\term}{\ty}}
    {\expjb{\SIG}{\etenv}{\ABS{\var}{\term}}{\ABSTY{\nty}{\ty}}} \and
    \inferrule{\expjb{\SIG}{\etenv}{\term}{\ty} \\ \expjb{\SIG}{\etenv}{\term'}{\ty}}
    {\formjb{\SIG}{\etenv}{\ECON{\term}{\term'}}} \and
    \inferrule{\expjb{\SIG}{\etenv}{\var}{\nty} \\ \expjb{\SIG}{\etenv}{\term}{\ty}}
    {\formjb{\SIG}{\etenv}{\FCON{\var}{\term}}}
  \end{mathpar}}
  \caption{Typing rules for non-permutative nominal terms and atomic constraints}
  \label{fig:typing}
\end{figure}

We let $\etenv$ range over typing environments, which are finite
partial functions from $\Var$ to $\Ty{\SIG}$ which assign types to variables.
We write $\dom{\etenv}$ for the domain of definition of $\etenv$.
\Fig{\ref{fig:typing}} provides rules which define typing
judgements of the form $\expjb{\SIG}{\etenv}{\term}{\ty}$ for terms
and $\formjb{\SIG}{\etenv}{\constr}$ for atomic constraints.
Note that if $\expjb{\SIG}{\etenv}{\term}{\nty}$ holds then
$\term$ must be a variable $\var$ such that $\etenv\pn{\var}=\nty$
for some name sort $\nty\in\SIGNtys{\SIG}$.

\begin{defi}[Non-permutative nominal constraint problems]
\label{def:cprob}
  Let $\prob$ range over finite conjunctions
  $\CONJRANGE{\constr_1}{\constr_k}$ of atomic constraints,
  where each $\constr_i\in\Constr{\SIG}$.
  Then, a non-permutative nominal constraint problem in $\Prob{\SIG}$
  has the form $\cprob{\etenv}{\prob}$, and is well-formed
  \pn{written $\formjb{\SIG}{\emptyset}{\cprob{\etenv}{\prob}}$}
  iff $\formjb{\SIG}{\etenv}{\constr}$ holds for all $\constr\in\prob$.
\end{defi}

Having specified the syntax of non-permutative terms and
constraint problems, we now turn to their semantics.
We will give a semantics to atomic constraints and
constraint problems in terms of instantiations
of their variables with \al-equivalence classes
of ground abstract syntax trees.
To do so, we must first define a notion of
ground trees quotiented by \al-equivalence.

\begin{defi}[\al-trees]
\label{def:al-trees}
  Let $\Atree{\SIG}{\ty}$ be the set of all 
  $\aeqsym$-equivalence classes of ground trees of type $\ty$,
  which we call \emph{\al-trees}.
  We let $\altree$ range over \al-trees.
  If $\tree\in\Tree{\SIG}{\ty}$ then we write $\aec{\tree}$ for the set
  $\setwhere{\tree'}{\aeq{\tree}{\tree'}{\ty}}$ of all ground trees
  which are \al-equivalent to $\tree$.
  If $\tree\in\Tree{\SIG}{\ty}$ then $\aec{\tree}\in\Atree{\SIG}{\ty}$.
  In the special case where $\tree\in\Atree{\SIG}{\nty}$
  it follows that $\tree=\aec{\name}=\setb{\name}$,
  for some $\name\in\Name\pn{\nty}$
  \pn{since constructors cannot produce trees of name sorts}.
\end{defi}

\al-trees represent the object-language terms quotiented by
\al-equivalence which are so frequently used in informal
mathmatical parlance.
They will form the basis for the semantics of
non-permutative nominal constraints.
We now extend the standard notions of \qtn{free names}
and \qtn{freshness} from nominal abstract syntax
to \al-trees.

\begin{defi}[Free names and freshness for \al-trees]
\label{def:fn-fresh}
  Suppose that $\altree\in\Atree{\SIG}{\ty}$.
  Then, we write $\fn{\altree}$ for the finite set
  $\fn{\tree}$ for some/any\footnote{Some/any
  properties are characteristic of nominal techniques for
  representing abstract syntax---see \cite{PittsAM:alpsri}
  for a rigorous mathematical treatment.} ground tree
  $\tree\in\Tree{\SIG}{\ty}$ such that $\altree=\aec{\tree}$.
  Furthermore, if $\altree\in\Atree{\SIG}{\nty}$ then we know that
  $\altree=\aec{\name}$ for some $\name\in\Name\pn{\nty}$.
  Then, we write $\freshfor{\altree}{\altree'}$ and say \qtn{$\altree$ is
  fresh for $\altree'$} iff $\name\notin\fn{\altree'}$.
\end{defi}

We now describe the instantiation of schematic terms,
which involves replacing the variables with \al-trees
to produce specific ground instances.

\begin{defi}[\al-tree valuations]
\label{def:vln}
  An \emph{\al-tree valuation} $\vln$ is a finite partial function which
  maps variables to \al-trees.
  We write $\dom{\vln}$ for the domain of the partial function.
  Given a type environment $\etenv$ we write $\Atree{\SIG}{\etenv}$ for the set of
  all \al-tree valuations $\vln$ such that $\dom{\vln}=\dom{\etenv}$ and
  $\vln\pn{\var}\in\Atree{\SIG}{\etenv\pn{\var}}$ for all $\var\in\dom{\vln}$.
  This ensures that the valuation respects types.
\end{defi}

Using the proof techniques from \cite{PittsAM:alpsri}
we can show that there exists an instantiation operation
$\vapp{\vln}{\term}$ which respects both types
and \al-equivalence classes, i.e.~if $\expjb{\SIG}{\etenv}{\term}{\ty}$
and $\vln\in\Atree{\SIG}{\etenv}$ then $\vapp{\vln}{\term}\in\Atree{\SIG}{\ty}$
holds, where $\vapp{\vln}{\term}$ is as follows.
\begin{align*}
  \vapp{\vln}{\var} &= \vln\pn{\var} \\
  \vapp{\vln}{\term}=\aec{\tree} \;\imp\; \vapp{\vln}{\APP{\CON}{\term}}
  &= \aec{\APP{\CON}{\tree}} \\
  \vapp{\vln}{\UNIT} &= \aec{\UNIT} \\
  \vapp{\vln}{\term_1}=\aec{\tree_1} \conj \cdots \conj \vapp{\vln}{\term_k}=\aec{\tree_k}
  \;\imp\; \vapp{\vln}{\TUPLE{\term_1}{\term_k}}
  &= \aec{\TUPLE{\tree_1}{\tree_k}} \\
  \vln\pn{\var}=\aec{\name} \conj \vapp{\vln}{\term}=\aec{\tree} \;\imp\;
  \vapp{\vln}{\ABS{\var}{\term}} &= \aec{\ABS{\name}{\tree}}
\end{align*}
Since variables stand for unknown \al-trees,
not unknown trees, we see that schematic terms
describe an \al-tree as opposed to a tree.
Precisely which one depends on how the variables
in the term are instantiated.
This reflects the common practice of leaving
\al-equivalence implicit and using representatives
to stand in place of the whole class
\cite[\Conv{2.1.13}]{BarendregtHP:lamcss}.

Applying an \al-tree valuation to a schematic term
is \qtn{possibly-capturing} with regard to the
binders in the underlying language of ground
abstract syntax trees, even when in abstraction position.
For example, given distinct variables $\vx$, $\vy$, $\vz$
we cannot regard the schematic terms $\ABS{\var}{\vz}$ and $\ABS{\vy}{\vz}$
as equivalent because if we let
$\vln=\setb{\var\mapsto\aec{\name},\vy\mapsto\aec{\name'},\vz\mapsto\aec{\name}}$
\pn{with $\name\neq\name'$} then we get
\begin{displaymath}
  \vapp{\vln}{\ABS{\var}{\vz}}=\aec{\ABS{\name}{\name}}\neq
  \aec{\ABS{\name'}{\name}}=\vapp{\vln}{\ABS{\vy}{\vz}}.
\end{displaymath}
Barendregt \cite{BarendregtHP:lamcss} does not draw a distinction
between names and schematic variables, but in our presentation
there is a clear distinction between non-permutative variables $\var$
and permutative names $\name$.
Even when in abstraction position, we allow non-permutative variables
to be \emph{aliased}, that is, we allow syntactically distinct variables
to be assigned with the same underlying name.
We can model names which behave permutatively by imposing additional
constraints that the variables must be mutually distinct
\cite[\Sect{6.4}]{LakinMR:exemli}.

\begin{lem}[Substitution lemma]
  \label{lem:sub-vln}
  If $\expjb{\SIG}{\etenv}{\term'}{\ty}$ and
  $\formjb{\SIG}{\etenv}{\ECON{\var}{\term}}$ and
  $\vln\in\Atree{\SIG}{\etenv}$ all hold
  then $\vln\pn{\var}=\vapp{\vln}{\term}$ implies
  $\vapp{\vln}{\term'}=\vapp{\vln}{\sapp{\sub{\var}{\term}}{\term'}}$.\qed
\end{lem}

We are now in a position to define the semantics of
non-permutative nominal constraint problems,
in terms of satisfying ground instantiations
by \al-tree valuations.

\begin{defi}[Satisfaction of atomic constraints]
\label{def:constr-sat}
  For an atomic constraint $\constr$ such that
  $\formjb{\SIG}{\etenv}{\constr}$ and an \al-tree valuation
  $\vln\in\Atree{\SIG}{\etenv}$, we write
  $\SATIS{\vln}{\constr}$ to mean that
  \qtn{$\vln$ satisfies $\constr$},
  which is defined by cases on $\constr$.
  \begin{mathpar}
    \inferrule{\vapp{\vln}{\term}=\vapp{\vln}{\term'}}
              {\SATIS{\vln}{\ECON{\term}{\term'}}} \and
    \inferrule{\freshfor{\vln\pn{\var}}{\vapp{\vln}{\term}}}
              {\SATIS{\vln}{\FCON{\var}{\term}}}
  \end{mathpar}
  In the case of freshness constraints, this relation is
  well-defined by virtue of the points noted in \Defn{\ref{def:fn-fresh}}.
\end{defi}

\begin{defi}[Satisfiable constraint problems]
\label{def:prob-sat}
  For a constraint problem $\cprob{\etenv}{\prob}$ such that
  $\formjb{\SIG}{\emptyset}{\cprob{\etenv}{\prob}}$, we say that
  $\SAT{\cprob{\etenv}{\prob}}$
  holds iff there exists a valuation $\vln\in\Atree{\SIG}{\etenv}$
  such that $\SATIS{\vln}{\constr}$ for all $\constr\in\prob$.
\end{defi}

For example, suppose that we have variables $\var$ and $\vy$ of some
name sort $\nty$.
Then, the \al-tree constraint problem
$\ECON{\ABS{\var}{\vy}}{\ABS{\vy}{\var}}$
is satisfied by any valuation $\vln$ such that
$\vln\pn{\var}=\vln\pn{\vy}$, as both sides of the equality constraint
are then instantiated to the same \al-equivalence class $\aec{\ABS{\name}{\name}}$.
Note that if we used permutative names $\name_{\var}$ and $\name_{\vy}$
then the corresponding constraint problem
$\ECON{\ABS{\name_{\var}}{\name_{\vy}}}{\ABS{\name_{\vy}}{\name_{\var}}}$
would \emph{not} be satisfiable, because the two terms are ground but
are not in the same \al-equivalence class.
This corresponds to the non-permutative constraint problem
$\CONJ{\ECON{\ABS{\var}{\vy}}{\ABS{\vy}{\var}}}{\FCON{\var}{\vy}}$
where we simulate permutative behaviour by adding appropriate
freshness constraints.
This constraint problem is also unsatisfiable, because the first constraint
$\ECON{\ABS{\var}{\vy}}{\ABS{\vy}{\var}}$ is only satisfiable
by a valuation $\vln$ if $\vln\pn{\var}=\vln\pn{\vy}$.
However, no such valuation can satisfy the freshness constraint
$\FCON{\var}{\vy}$.

\begin{defi}
\label{def:aml-sat-nomsat}
  We write \NonPermSat{} for the decision problem
  $\setwhere{\pair{\etenv}{\prob}}{\SAT{\cprob{\etenv}{\prob}}}$.
\end{defi}

It is trivial to see that \NonPermSat{} is decidable because it is a
syntactic subset of the equivariant unification problem,
which was shown to be decidable by Cheney in his thesis
\cite[\Chap{7}]{CheneyJ:nomlp-phd}.
In previous work \cite{LakinMR:residb} we showed that \NonPermSat{}
is is \NP{}-hard by a reduction of graph 3-colourability.
In \Sect{\ref{sec:equivariant}} of this paper we present an alternative
proof of \NP{}-hardness by defining an encoding of name-name
equivariant unification problems \pn{which are known to be
\NP{}-complete \cite{CheneyJ:equu-jv}} into NPNAS.
It follows from \cite[\Thm{1}]{CheneyJ:equu-jv} that
\NonPermSat{} is in \NP{} and hence that \NonPermSat{}
is \NP{}-complete.

\begin{rem}[Permutations and equivariance]
\label{rem:permutations}
  This section has hardly mentioned name-permutations,
  which are a staple of most nominal techniques for
  abstract syntax involving binders
  \pn{as discussed in \Sect{\ref{subsec:background-nas}}}.
  They are not required because \al-equivalence
  is handled by the explicit use of \al-equivalence
  classes of ground trees in the semantics.
  However, it is not hard to show that the semantics
  of non-permutative nominal constraints is equivariant,
  that is, that \pn{for any $\constr$} the set
  $\setwhere{\vln}{\SATIS{\vln}{\constr}}$ is
  closed under name-permutations.
\end{rem}

\section{Constraint transformation algorithm}
\label{sec:trans}

We define a non-deterministic transition relation, ${\trans}$,
which transforms a single constraint problem into a
finite, non-empty set of constraint problems.
\Fig{\ref{fig:trans}} presents transition rules for the
transformation relation.
To save space, we write $\fcons{\var}{\var}{1}{k}$
for the conjunction $\CONJRANGE{\FCON{\var}{\var_1}}{\FCON{\var}{\var_k}}$.
We also write $\abss{\var}{1}{k}{\term}$ as a shorthand for the iterated
abstraction term $\ABS{{\var}_{1}}{\cdots\ABS{{\var}_{k}}{\term}}$,
where the list of abstractions may be empty unless explicitly stated otherwise.
Note that \Fig{\ref{fig:trans}} does not contain explicit rules for handling
constraints of the form
$\ECON{\abss{\var}{1}{k}{\ABS{\var}{\term}}}{\abss{\vy}{1}{k}{\ABS{\vy}{\term'}}}$
or $\FCON{\var}{\abss{\vy}{1}{k}{\ABS{\vy}{\term'}}}$ because the
additional abstraction term-former is implicitly folded into the
initial abstraction list during pattern-matching.

Rules \frule{1}--\frule{3} and \erule{1}--\erule{3} deal with
unit, data and tuple terms in the usual way:
the only difference is that we work within nested abstractions.
The abstractions do not play any part in these six rules,
except that the lists on both sides of an equality constraint
must be of the same length.
The rules \frule{5} and \erule{5} dispose of trivial constraints:
in the case of \frule{5}, two names of different sorts will
always be fresh for each other and in the case of \erule{5},
any term is equal to itself.

The most interesting rules are \frule{4} and \erule{4}, which deal with
the scopes of bound names with respect to the nested abstractions.
We first consider \frule{4}.
In order for $\var$ to not appear free anywhere in
${\abss{\vy}{1}{k}{\vy}}$, either $\var$ should map to the same name
as one of the abstracted variables $\lrange{\vy_1}{\vy_k}$
or $\var$ should be distinct from all of the abstracted variables
and be constrained to be fresh for the unknown term $\vy$.
Unlike in nominal unification, transforming a freshness constraint
with this rule may produce new equality constraints to solve.
 
Rule \erule{4} deals with equality constraints between variables
of some name sort $\nty$.
We handle these constraints by noting that the way to resolve the
binding scope of the names $\var$ and $\vy$ is to start at the
innermost binding occurrence and work towards the outside.
Therefore, it should be the case that either $\var$ and $\vy$ both unify with the
innermost binder \pn{$\var_k$ and $\vy_k$ respectively}, or that they should both be
distinct from the innermost binder and unify with the next one moving outwards
\pn{i.e.~$\var_{k-1}$ and $\vy_{k-1}$}, and so on, or that $\var$ and $\vy$
should be distinct from all of the potential binders and equal to each other.
This method of dealing with equality constraints between bound names
seems more natural than existing methods based on name-swapping.

The final two rules, \erule{6} and \erule{7}, eliminate variables from the problem
by substituting throughout the remaining constraints.
They use a notion of substitution $\sapp{\sub{\var}{\term}}{\prob}$
which replaces all occurrences of the variable $\var$ in $\prob$
by the term $\term$.
These substitutions are capturing with respect to the
abstraction term-former.
Rule \erule{6} is the standard variable elimination rule from first-order
\pn{syntactic} unification.
The side-condition $\var\notin\vars{\term}$ on this rule enforces the
\emph{occurs check} which is necessary to avoid cyclic substitutions.
The side-condition $\var\in\vars{\prob}$ ensures that this rule can
only be invoked once per variable, which is necessary for termination.

Rule \erule{7} deals with equality constraints of the form
$\ECON{\abss{\var}{1}{k}{\var}}{\abss{\vy}{1}{k}{\term}}$,
where $k>0$, the occurs check succeeds and $\term$ is not a variable,
i.e.~$\term$ is some compound term.
This includes the case where there are more initial nested abstractions
on one side than the other.
We cannot simply substitute $\term$ for $\var$ here because
of the preceding abstractions: $\var$ might
need to be instantiated with a term syntactically different from $\term$.
For example, to satisfy the constraints
$\CONJ{\pn{\FCON{\var}{\vy}}}
{\pn{\ECON{\ABS{\var}{\twinkle{\var}}}{\ABS{\vy}{\APP{\CON}{\vy}}}}}$
it is clear that $\twinkle{\var}$ must be mapped to
$\APP{\CON}{\var}$, not $\APP{\CON}{\vy}$.
This is where swappings are necessary in both nominal and equivariant
unification: however, this is not an elegant solution when bound names are
represented using variables, because the potential for aliasing means that
the result of a \qtn{variable swapping} such as $\pact{\swap{\var}{\vy}}{\vz}$
is not unique.

Since we cannot make progress using a swapping, we note that the
side-condition that $\term$ may not be a variable means that we know
the outermost constructor of $\term$.  This allows us to impose some
structure on the unknown term represented by $\var$ by narrowing
\cite{AntoyS:neens}.  The rules from \Fig{\ref{fig:narrow}} define a
narrowing relation which factors out this common functionality at
unit, tuple, data and abstraction types.

\begin{figure}
  \begin{desCription}
    \item\noindent{\hskip-12 pt\bf \frule{1}:}\
    $\cprob{\etenv}{\CONJ{\pn{\FCON{\var}{\abss{\vy}{1}{k}{\UNIT}}}}{\prob}}
    \trans \cprob{\etenv}{\prob}$
    \item\noindent{\hskip-12 pt\bf \frule{2}:}\
    $\cprob{\etenv}
    {\CONJ{\pn{\FCON{\var}{\abss{\vy}{1}{k}{\APP{\CON}{\term}}}}}{\prob}}
    \trans \cprob{\etenv}
    {\CONJ{\pn{\FCON{\var}{\abss{\vy}{1}{k}{\term}}}}{\prob}}$
    \item\noindent{\hskip-12 pt\bf \frule{3}:}\
    $\cprob{\etenv}
    {\CONJ{\pn{\FCON{\var}{\abss{\vy}{1}{k}{\TUPLE{\term_1}{\term_j}}}}}{\prob}}$ \\
    $\trans \cprob{\etenv}{\CONJ{\CONJRANGE
    {\pn{\FCON{\var}{\abss{\vy}{1}{k}{\term_1}}}}
    {\pn{\FCON{\var}{\abss{\vy}{1}{k}{\term_j}}}}}{\prob}}$
    \item\noindent{\hskip-12 pt\bf \frule{4}:}\
    $\cprob{\etenv}{\CONJ{\pn{\FCON{\var}{\abss{\vy}{1}{k}{\vy}}}}{\prob}}$ \\
    $\trans \left\{
    \begin{array}{l@{\quad}l}
      \cprob{\etenv}{\CONJ{\pn{\ECON{\var}{\vy_1}}}{\prob}}
      & \text{if $\etenv\pn{\var}=\etenv\pn{\vy_1}$.} \\
      \cprob{\etenv}{\CONJ{\pn{\FCON{\var}{\vy_1}}}
      {\CONJ{\pn{\ECON{\var}{\vy_2}}}{\prob}}}
      & \text{if $\etenv\pn{\var}=\etenv\pn{\vy_2}$.} \\
      \cdots & \cdots \\
      \cprob{\etenv}{\CONJ{\pn{\fcons{\var}{\vy}{1}{k-1}}}
      {\CONJ{\pn{\ECON{\var}{\vy_k}}}{\prob}}}
      & \text{if $\etenv\pn{\var}=\etenv\pn{\vy_k}$.} \\
      \cprob{\etenv}{\CONJ{\pn{\fcons{\var}{\vy}{1}{k}}}
      {\CONJ{\pn{\FCON{\var}{\vy}}}{\prob}}}
    \end{array}\right.$ \\
    if $k>0$.
    \item\noindent{\hskip-12 pt\bf \frule{5}:}\
    $\cprob{\etenv}{\CONJ{\pn{\FCON{\var}{\vy}}}{\prob}}
    \trans \cprob{\etenv}{\prob}$ \\
    if $\etenv\pn{\var}=\nty$, $\etenv\pn{\vy}=\nty'$ and $\nty\neq\nty'$.
    \item\noindent{\hskip-12 pt\bf \erule{1}:}\ $\cprob{\etenv}
    {\CONJ{\pn{\ECON{\abss{\var}{1}{k}{\UNIT}}{\abss{\vy}{1}{k}{\UNIT}}}}{\prob}}
    \trans \cprob{\etenv}{\prob}$
    \item\noindent{\hskip-12 pt\bf \erule{2}:}\ $\cprob{\etenv}
    {\CONJ{\pn{\ECON{\abss{\var}{1}{k}{\APP{\CON}{\term}}}
    {\abss{\vy}{1}{k}{\APP{\CON}{\term'}}}}}{\prob}}
    \trans \cprob{\etenv}
    {\CONJ{\pn{\ECON{\abss{\var}{1}{k}{\term}}{\abss{\vy}{1}{k}{\term'}}}}{\prob}}$
    \item\noindent{\hskip-12 pt\bf \erule{3}:}\ $\cprob{\etenv}
    {\CONJ{\pn{\ECON{\abss{\var}{1}{k}{\TUPLE{\term_1}{\term_j}}}
    {\abss{\vy}{1}{k}{\TUPLE{\term_1'}{\term_j'}}}}}{\prob}}$ \\
    $\trans \cprob{\etenv}
    {\CONJ{\CONJRANGE{\pn{\ECON{\abss{\var}{1}{k}{\term_1}}{\abss{\vy}{1}{k}{\term_1'}}}}
    {\pn{\ECON{\abss{\var}{1}{k}{\term_j}}{\abss{\vy}{1}{k}{\term_j'}}}}}{\prob}}$
    \item\noindent{\hskip-12 pt\bf \erule{4}:}\ $\cprob{\etenv}
    {\CONJ{\pn{\ECON{\abss{\var}{1}{k}{\var}}{\abss{\vy}{1}{k}{\vy}}}}{\prob}}$ \\
    $\trans\left\{
    \begin{array}{l@{\quad}l}
      \cprob{\etenv}{\CONJ{\pn{\ECON{\var}{\var_k}}}
      {\CONJ{\pn{\ECON{\vy}{\vy_k}}}{\prob}}}
      & \text{if $\etenv\pn{\var}=\etenv\pn{\var_k}$.} \\
      \cprob{\etenv}{\CONJ{\pn{\FCON{\var}{\var_k}}}
      {\CONJ{\pn{\ECON{\var}{\var_{k-1}}}}
      {\CONJ{\pn{\FCON{\vy}{\vy_k}}}
      {\CONJ{\pn{\ECON{\vy}{\vy_{k-1}}}}{\prob}}}}}
      & \text{if $\etenv\pn{\var}=\etenv\pn{\var_{k-1}}$.} \\
      \cdots & \cdots \\
      \cprob{\etenv}{\CONJ{\pn{\fcons{\var}{\var}{k}{2}}}
      {\CONJ{\pn{\ECON{\var}{\var_1}}}
      {\CONJ{\pn{\fcons{\vy}{\vy}{k}{2}}}
      {\CONJ{\pn{\ECON{\vy}{\vy_1}}}{\prob}}}}}
      & \text{if $\etenv\pn{\var}=\etenv\pn{\var_1}$.} \\
      \cprob{\etenv}{\CONJ{\pn{\fcons{\var}{\var}{k}{1}}}
      {\CONJ{\pn{\fcons{\vy}{\vy}{k}{1}}}
      {\CONJ{\pn{\ECON{\var}{\vy}}}{\prob}}}}
    \end{array}\right.$ \\
    if $k>0$ and $\etenv\pn{\var}=\nty$, for some $\nty$.
    \item\noindent{\hskip-12 pt\bf \erule{5}:}\ $\cprob{\etenv}{\CONJ{\pn{\ECON{\var}{\var}}}{\prob}}
    \trans \cprob{\etenv}{\prob}$
    \item\noindent{\hskip-12 pt\bf \erule{6}:}\
    $\left.\begin{array}{l}
      \cprob{\etenv}{\CONJ{\pn{\ECON{\var}{\term}}}{\prob}} \\
      \cprob{\etenv}{\CONJ{\pn{\ECON{\term}{\var}}}{\prob}}
    \end{array}\right\}\trans
    \cprob{\etenv}{\CONJ{\pn{\ECON{\var}{\term}}}
    {\sapp{\sub{\var}{\term}}{\prob}}}$ \\
    if $\var\notin\vars{\term}$ and $\var\in\vars{\prob}$.
    \item\noindent{\hskip-12 pt\bf \erule{7}:}\
    $\left.\begin{array}{l}
      \cprob{\etenv}
      {\CONJ{\pn{\ECON{\abss{\var}{1}{k}{\var}}
      {\abss{\vy}{1}{k}{\term}}}}{\prob}} \\
      \cprob{\etenv}
      {\CONJ{\pn{\ECON{\abss{\vy}{1}{k}{\term}}
      {\abss{\var}{1}{k}{\var}}}}{\prob}}
    \end{array}\right\}$ \\
    $\trans\cprob{\etenv,\twinkle{\etenv}}{\CONJ{\pn{\ECON{\var}{\twinkle{\term}}}}
    {\CONJ{\pn{\ECON{\abss{\var}{1}{k}{\twinkle{\term}}}{\abss{\vy}{1}{k}{\term}}}}
    {\sapp{\sub{\var}{\twinkle{\term}}}{\prob}}}}$ \\
    if $\term$ is not a variable, $\var\notin\vars{\term}$,
    $k>0$ and $\narrow{\etenv}{\term}{\twinkle{\etenv}}{\twinkle{\term}}$.
  \end{desCription}
  \caption{Constraint transformation rules}
  \label{fig:trans}
\end{figure}

\begin{figure}
  \mbox{\begin{mathpar}
    \inferrule*[left=\nrule{1}]
    {\ }{\narrow{\etenv}{\UNIT}{\emptyset}{\UNIT}} \and
    \inferrule*[left=\nrule{2}]
    {\var\notin\dom{\etenv} \\ \CONTYPE{\CON}{\ty}{\dty}{\SIG}}
    {\narrow{\etenv}{\APP{\CON}{\term}}{\setb{\var\OFTY\ty}}{\APP{\CON}{\var}}} \and
    \inferrule*[left=\nrule{3}]
    {\expjb{\SIG}{\etenv}{\TUPLE{\term_1}{\term_k}}{\TUPTY{\ty_1}{\ty_k}} \\
    \var_1\neq\cdots\neq\var_k\notin\dom{\etenv}}
    {\narrow{\etenv}{\TUPLE{\term_1}{\term_k}}
    {\setrange{\var_1\OFTY\ty_1}{\var_k\OFTY\ty_k}}
    {\TUPLE{\var_1}{\var_k}}} \and
    \inferrule*[left=\nrule{4}]
    {\expjb{\SIG}{\etenv}{\ABS{\var}{\term}}{\ABSTY{\nty}{\ty}} \\
    \var'\neq\var''\notin\dom{\etenv}}
    {\narrow{\etenv}{\ABS{\var}{\term}}
    {\setb{\var'\OFTY\nty,\var''\OFTY\ty}}{\ABS{\var'}{\var''}}}
  \end{mathpar}}
  \caption{Narrowing rules}
  \label{fig:narrow}
\end{figure}

The intuitive reading of $\narrow{\etenv}{\term}{\twinkle{\etenv}}{\twinkle{\term}}$
is that the term $\twinkle{\term}$ represents a \qtn{pattern} for terms with
the same outermost constructor as $\term$.
The subterms of $\twinkle{\term}$ are variables which stand for the \pn{as-yet unknown}
subterms of the term referred to by the variable $\var$.
The extra type environment $\twinkle{\etenv}$ is needed to ensure that the variables used to
create $\twinkle{\term}$ do not appear elsewhere in the constraint problem.
They must also be mutually distinct, as in rules \nrule{3} and \nrule{4}.
This gives rise to the following lemma, which is proved by cases on the
narrowing rules.

\begin{lem}[Narrowing and typing]
  \label{lem:narrowing-typing}
  If $\narrow{\etenv}{\term}{\twinkle{\etenv}}{\twinkle{\term}}$ and
  $\expjb{\SIG}{\etenv}{\term}{\ty}$ then
  $\dom{\etenv}\inter\dom{\twinkle{\etenv}}=\emptyset$ and
  $\expjb{\SIG}{\twinkle{\etenv}}{\twinkle{\term}}{\ty}$.\qed
\end{lem}

The narrowing procedure is lazy in the sense that each
narrowing step using rule \erule{7} does not replicate the entire
structure of the term $\term$ but just its outermost constructor.
If there is further structure on one side, rule \erule{7}
may be applied repeatedly.
There is no rule for narrowing against variables
because they have no internal structure to copy.
Constraints of the form
$\ECON{\abss{\var}{1}{k}{\var}}{\abss{\vy}{1}{k}{\vy}}$
are simply left alone when $\etenv\pn{\var}$ is not a
name sort---this is in direct contrast to nominal unification.
It is not immediately obvious that this is correct,
and we will address this point in the proof of
\Lem{\ref{lem:constraint-terminalsat}} below.

\begin{rem}[Relationship to existing algorithms]
\label{rem:existing-algorithms}
  Since \NonPermSat{} is a syntactic subset of the full equivariant
  unification problem studied by Cheney, the problems considered here
  could be handled using a subset of the rules presented in
  \cite{CheneyJ:equu-jv} for solving \qtn{general nominal unification} problems.
  The main difference is that our rules use the lazy narrowing approach
  to delay case analysis until the body of the abstractions is itself
  a variable of name sort.
  We also note that a simplified narrowing-based approach was used to solve
  swapping-free equivariant matching problems in polynomial time
  in \cite[\Sect{5.2}]{CheneyJ:equu-jv}.
\end{rem}

We conclude this section with the straightforward result
that well-formedness of constraint problems is preserved by
the transformation rules.
The proof is by cases on the transformation rules from
\Fig{\ref{fig:trans}}.
In the case for rule \erule{7} we require \Lem{\ref{lem:narrowing-typing}}
to deduce that the narrowing step preserves well-formedness.

\begin{lem}[Preservation of well-formedness]
  \label{lem:constraint-typres}
  If $\formjb{\SIG}{\emptyset}{\cprob{\etenv}{\prob}}$ and
  $\cprob{\etenv}{\prob}\trans\cprob{\etenv'}{\prob'}$ then
  $\formjb{\SIG}{\emptyset}{\cprob{\etenv'}{\prob'}}$.
  Furthermore, $\etenv'\supseteq\etenv$.\qed
\end{lem}

\section{A correct decision procedure}
\label{sec:correctness}

We now present an algorithm for deciding satisfiability
of non-permutative nominal constraint problems.
We begin by considering the correctness of
individual transformation rules from the
previous section and prove that,
with careful consideration of termination,
the transformation rules can be used
to give a correct decision procedure for \NonPermSat{}.

\subsection{Soundness and completeness of transformations}
\label{subsec:correctness-transformations}

We first prove soundness and completeness results for the
individual constraint transformation rules from
\Fig{\ref{fig:trans}}.
We begin by stating a lemma which relates substitution and
constraint satisfaction, which will be needed for the
cases for rules \erule{6} and \erule{7} which involve substitution.

\begin{lem}[Substitution property of satisfaction]
  \label{lem:constraint-satsubst}
  Suppose that
  $\formjb{\SIG}{\emptyset}{\cprob{\etenv,\var\OFTY\ty}{\prob}}$,
  $\expjb{\SIG}{\etenv}{\term}{\ty}$,
  $\vln\in\Atree{\SIG}{\etenv,\var\OFTY\ty}$ and
  $\vln\pn{\var}=\vapp{\vln}{\term}$.
  Then $\SATIS{\vln}{\sapp{\sub{\var}{\term}}{\prob}}$
  iff $\SATIS{\vln}{\prob}$.\qed
\end{lem}

We now prove that the transformation rules are \emph{sound},
i.e.~that the transformation steps do not introduce any
additional satisfying valuations to the problem.

\begin{thm}[Soundness of transformations]
  \label{thm:constraint-sound}
  Suppose that $\formjb{\SIG}{\emptyset}{\cprob{\etenv}{\prob}}$,
  $\cprob{\etenv}{\prob}\trans\cprob{\etenv'}{\prob'}$ and
  $\SATIS{\vln'}{\prob'}$ all hold,
  where $\vln'\in\Atree{\SIG}{\etenv'}$.
  Then $\SATIS{\vln}{\prob}$ holds, where $\vln$ is the restriction
  of $\vln'$ to $\dom{\etenv}$.
\end{thm}
\proof
  By case analysis on the transformation rule used to derive
  $\cprob{\etenv}{\prob}\trans\cprob{\etenv'}{\prob'}$.
  The cases for rules \frule{1}--\frule{4} and \erule{1}--\erule{4}
  are straightforward, using standard facts about the definition
  of constraint satisfaction.
  The case for \erule{5} follows because $\SATIS{\vln}{\ECON{\var}{\var}}$
  holds for any $\vln$ and $\var$.
  Similarly, the case for \frule{5} follows because
  $\SATIS{\vln}{\FCON{\var}{\vy}}$ holds for any $\vln$, $\var$ and $\vy$
  if $\etenv\pn{\var}$ and $\etenv\pn{\vy}$ are different name sorts.
  The case for \erule{6} relies on \Lem{\ref{lem:constraint-satsubst}}.
  The remaining case, for \erule{7}, is dealt with in detail below.
  \begin{desCription}

    \item\noindent{\hskip-12 pt\bf \erule{7}:}\
      In this case we have
      $\prob=\CONJ{\pn{\ECON{\abss{\var}{1}{k}{\var}}
      {\abss{\vy}{1}{k}{\term}}}}{\twinkle{\prob}}$ and furthermore that
      $\prob'=\CONJ{\pn{\ECON{\var}{\twinkle{\term}}}}
      {\CONJ{\pn{\ECON{\abss{\var}{1}{k}{\twinkle{\term}}}
      {\abss{\vy}{1}{k}{\term}}}}{\sapp{\sub{\var}{\twinkle{\term}}}{\prob}}}$,
      where $\term$ is not a variable, $\var\notin\vars{\term}$, $k>0$
      and $\narrow{\etenv}{\term}{\twinkle{\etenv}}{\twinkle{\term}}$.
      Furthermore, $\etenv'=\etenv,\twinkle{\etenv}$.
      By assumption we get that
      $\SATIS{\vln'}{\ECON{\var}{\twinkle{\term}}}$,
      $\SATIS{\vln'}{\ECON{\abss{\var}{1}{k}{\twinkle{\term}}}
      {\abss{\vy}{1}{k}{\term}}}$ and
      $\SATIS{\vln'}{\sapp{\sub{\var}{\twinkle{\term}}}{\twinkle{\prob}}}$
      all hold, for some $\vln'\in\Atree{\SIG}{\etenv,\twinkle{\etenv}}$.
      From $\SATIS{\vln'}{\ECON{\var}{\twinkle{\term}}}$
      we know that $\vln'\pn{\var}=\vapp{\vln'}{\twinkle{\term}}$,
      \pn{since $\var\notin\vars{\term}$}
      and then by \Lem{\ref{lem:constraint-satsubst}} and
      $\SATIS{\vln'}{\sapp{\sub{\var}{\twinkle{\term}}}{\twinkle{\prob}}}$
      we get that $\SATIS{\vln'}{\twinkle{\prob}}$ holds.
      Furthermore, we know that
      $\ECON{\abss{\var}{1}{k}{\twinkle{\term}}}{\abss{\vy}{1}{k}{\term}}$
      is $\sapp{\sub{\var}{\twinkle{\term}}}
      {\pn{\ECON{\abss{\var}{1}{k}{\var}}{\abss{\vy}{1}{k}{\term}}}}$,
      because $\var\notin\vars{\term}$.
      Therefore, by $\SATIS{\vln'}{\ECON{\abss{\var}{1}{k}{\twinkle{\term}}}
      {\abss{\vy}{1}{k}{\term}}}$ and \Lem{\ref{lem:constraint-satsubst}}
      we can show that $\SATIS{\vln'}{\ECON{\abss{\var}{1}{k}{\var}}
      {\abss{\vy}{1}{k}{\term}}}$ holds.
      Thus we get that $\SATIS{\vln'}{\prob}$ holds,
      and hence that $\SATIS{\vln}{\prob}$, as required.\qed

  \end{desCription}

\noindent Next we prove that the constraint transformation rules are
\emph{complete}, i.e.~that every satisfying valuation is preserved across
some transformation of a constraint problem.
We present some preliminary lemmas concerning narrowing before moving
on to the main proof.

\begin{lem}[Possibility of narrowing]
  \label{lem:narrowing-possibility}
  If $\term$ is not a variable and $\expjb{\SIG}{\etenv}{\term}{\ty}$
  then there exist $\twinkle{\etenv}$ and $\twinkle{\term}$ such that
  $\narrow{\etenv}{\term}{\twinkle{\etenv}}{\twinkle{\term}}$ holds.
\end{lem}
\proof
  By cases on the narrowing rules from \Fig{\ref{fig:narrow}},
  which cover all syntactic cases for $\term$ except for when
  $\term$ is a variable.
  In each case, the types from the typing assumption
  $\expjb{\SIG}{\etenv}{\term}{\ty}$ match those required by the
  appropriate narrowing rule, so we can apply that rule to
  find $\twinkle{\etenv}$ and $\twinkle{\term}$ such that
  $\narrow{\etenv}{\term}{\twinkle{\etenv}}{\twinkle{\term}}$,
  as required.\qed

\begin{lem}[Narrowing and satisfaction]
  \label{lem:narrowing-satisfaction}
  Suppose that $\formjb{\SIG}{\etenv}{\ECON{\abss{\var}{1}{k}{\var}}
  {\abss{\vy}{1}{k}{\term}}}$ and that $\vln\in\Atree{\SIG}{\etenv}$
  is such that $\SATIS{\vln}{\ECON{\abss{\var}{1}{k}{\var}}
  {\abss{\vy}{1}{k}{\term}}}$ holds.
  If $\narrow{\etenv}{\term}{\twinkle{\etenv}}{\twinkle{\term}}$
  then there exists $\twinkle{\vln}\in\Atree{\SIG}{\etenv,\twinkle{\etenv}}$
  which agrees with $\vln$ on $\dom{\etenv}$ and is such that
  $\twinkle{\vln}\pn{\var}=\vapp{\twinkle{\vln}}{\twinkle{\term}}$.
\end{lem}
\proof
  From $\narrow{\etenv}{\term}{\twinkle{\etenv}}{\twinkle{\term}}$ and
  \Lem{\ref{lem:narrowing-typing}} we get that
  $\dom{\etenv}\inter\dom{\twinkle{\etenv}}=\emptyset$.
  Thus it follows that there exist a family of valuations
  $\twinkle{\vln}\in\Atree{\SIG}{\etenv,\twinkle{\etenv}}$
  which agree with $\vln$ on $\dom{\etenv}$, and it just remains to show that
  $\twinkle{\vln}\pn{\var}=\vapp{\twinkle{\vln}}{\twinkle{\term}}$
  holds for some such $\twinkle{\vln}$.
  We prove this by cases on the narrowing rule used to derive
  $\narrow{\etenv}{\term}{\twinkle{\etenv}}{\twinkle{\term}}$.
  In each case, $\SATIS{\vln}{\ECON{\abss{\var}{1}{k}{\var}}
  {\abss{\vy}{1}{k}{\term}}}$ implies that the outermost
  term-former of $\twinkle{\vln}\pn{\var}$ is the same as that of
  $\vapp{\twinkle{\vln}}{\term}$, and since
  $\narrow{\etenv}{\term}{\twinkle{\etenv}}{\twinkle{\term}}$
  this is also the same as the outermost term-former of
  $\vapp{\twinkle{\vln}}{\twinkle{\term}}$.
  Thus we can choose a valuation $\twinkle{\vln}$ which instantiates
  the variables in $\dom{\twinkle{\etenv}}$ such that
  $\twinkle{\vln}\pn{\var}=\vapp{\twinkle{\vln}}{\twinkle{\term}}$
  holds, as required.\qed

\begin{defi}[Successor sets]
  \label{def:constraint-succ}
  We write $\succ{\cprob{\etenv}{\prob}}$ for the \emph{successor set} of
  $\cprob{\etenv}{\prob}$, which we define as the set
  $\setwhere{\cprob{\etenv'}{\prob'}}
  {\cprob{\etenv}{\prob}\trans\cprob{\etenv'}{\prob'}}$.
\end{defi}

\begin{thm}[Completeness of transformations]
  \label{thm:constraint-complete}
  Suppose that $\formjb{\SIG}{\emptyset}{\cprob{\etenv}{\prob}}$,
  $\vln\in\Atree{\SIG}{\etenv}$ and $\SATIS{\vln}{\prob}$ all hold,
  and that $\succ{\cprob{\etenv}{\prob}}\neq\emptyset$.
  Then there exists $\cprob{\etenv'}{\prob'}\in\succ{\prob}$
  and $\vln'\in\Atree{\SIG}{\etenv'}$ such that $\SATIS{\vln'}{\prob'}$,
  where $\vln$ and $\vln'$ agree on $\dom{\etenv}$.
\end{thm}
\proof
  Since $\succ{\cprob{\etenv}{\prob}}\neq\emptyset$ it follows that
  $\prob$ matches the left-hand side of one of the constraint transformation
  rules from \Fig{\ref{fig:trans}} and satisfies any side-conditions.
  Then, the proof is by case analysis on $\prob$.
  The cases of $\prob$ which match rules \frule{1}--\frule{4} and
  \erule{1}--\erule{4} are straightforward and follow from
  standard properties of constraint satisfaction.

  If $\prob$ is $\CONJ{\pn{\ECON{\var}{\var}}}{\prob}$ the transition
  uses rule \erule{5} and the result is trivial by assumption,
  the problem on the right-hand side being a subset of the problem
  on the left-hand side.
  If $\prob$ is $\CONJ{\pn{\FCON{\var}{\vy}}}{\prob}$, where
  $\var$ and $\vy$ are distinct variables of different name sorts,
  then \frule{5} applies and again the result follows trivially.
  We invoke \Lem{\ref{lem:constraint-satsubst}} in the case for \erule{6}.
  The case for \erule{7} is less straightforward, and we give details
  for this below.
  \begin{desCription}

    \item\noindent{\hskip-12 pt\bf Case $\bm{\prob = \CONJ{\pn{\ECON
    {\abss{\var}{1}{k}{\var}}{\abss{\vy}{1}{k}{\term}}}}{\twinkle{\prob}}}$:}\
      We also assume that $\term$ is not a variable, $\var\notin\vars{\term}$,
      and $k>0$ all hold.
      Furthermore, we assume that $\vln\in\Atree{\SIG}{\etenv}$, where
      $\SATIS{\vln}{\ECON{\abss{\var}{1}{k}{\var}}{\abss{\vy}{1}{k}{\term}}}$
      and $\SATIS{\vln}{\twinkle{\prob}}$ both hold.
      Since $\term$ is not a variable, by \Lem{\ref{lem:narrowing-possibility}}
      we get that $\narrow{\etenv}{\term}{\twinkle{\etenv}}{\twinkle{\term}}$ holds for some
      $\twinkle{\etenv}$ and $\twinkle{\term}$.
      We can then match against rule \erule{7} and get that
      $\prob'=\CONJ{\pn{\ECON{\var}{\twinkle{\term}}}}
      {\CONJ{\pn{\ECON{\abss{\var}{1}{k}{\twinkle{\term}}}{\abss{\vy}{1}{k}{\term}}}}
      {\sapp{\sub{\var}{\twinkle{\term}}}{\twinkle{\prob}}}}$
      and $\etenv'=\etenv,\twinkle{\etenv}$.
      By \Lem{\ref{lem:narrowing-satisfaction}} there exists
      $\vln'\in\Atree{\SIG}{\etenv'}$ which agrees with
      $\vln$ on $\dom{\etenv}$ and is such that
      $\vln'\pn{\var}=\vapp{\vln'}{\twinkle{\term}}$.
      It follows that $\SATIS{\vln'}{\ECON{\var}{\twinkle{\term}}}$
      and $\SATIS{\vln'}{\ECON{\abss{\var}{1}{k}{\twinkle{\term}}}
      {\abss{\vy}{1}{k}{\term}}}$ both hold.
      Finally, we can use \Lem{\ref{lem:constraint-satsubst}} to deduce that
      $\SATIS{\vln'}{\sapp{\sub{\var}{\twinkle{\term}}}{\twinkle{\prob}}}$,
      and hence that $\SATIS{\vln'}{\prob'}$ holds, as required.\qed

  \end{desCription}

\noindent The following corollary follows immediately from
\Thm{\ref{thm:constraint-sound}} and
\Thm{\ref{thm:constraint-complete}},
and summarises the results of this section.

\begin{cor}[Soundness and completeness of transformations]
  \label{cor:constraint-sndcmp}
  Assume $\formjb{\SIG}{\emptyset}{\cprob{\etenv}{\prob}}$
  and $\succ{\cprob{\etenv}{\prob}}=
  \setrange{\cprob{\etenv,\etenv_1}{\prob_1}}
  {\cprob{\etenv,\etenv_k}{\prob_k}}$ both hold.
  Then, for any $\vln\in\Atree{\SIG}{\etenv}$,
  $\SATIS{\vln}{\prob}$ iff there exists a valuation
  $\vln'$ which extends $\vln$ to $\dom{\etenv,\etenv_i}$
  and is such that $\SATIS{\vln'}{\prob_i}$,
  for some $i\in\setrange{1}{k}$.\qed
\end{cor}

\subsection{Termination}
\label{subsec:correctness-termination}

A key feature of any decision procedure is that it must
always terminate, but here we run into a problem:
for some constraint problems it is possible to
get infinite reduction sequences.
For example, given a datatype $\Natdty$ for
natural numbers, if $n>0$ then the constraint problem
\begin{equation*}
  \cprob{\etenv}{\CONJ
  {\pn{\ECON{\abss{\var}{1}{k}{\var}}{\abss{\vy}{1}{k}{\APP{\SKW}{\vy}}}}}
  {\pn{\ECON{\abss{\vy}{1}{k}{\vy}}{\abss{\var}{1}{k}{\APP{\SKW}{\var}}}}}}
\end{equation*}
can be reduced to
\begin{displaymath}
  \cprob{\etenv,\var'\OFTY\Natdty,\vy'\OFTY\Natdty}{\CONJ
  {\pn{\ECON{\var}{\APP{\SKW}{\var'}}}}{\CONJ
  {\pn{\ECON{\vy}{\APP{\kw{S}}{\vy'}}}}{\CONJ
  {\pn{\ECON{\abss{\var}{1}{k}{\var'}}{\abss{\vy}{1}{k}{\APP{\kw{S}}{\vy'}}}}}
  {\pn{\ECON{\abss{\vy}{1}{k}{\vy'}}{\abss{\var}{1}{k}{\APP{\kw{S}}{\var'}}}}}}}}.
\end{displaymath}
There is clearly the possibility of divergence as we have recovered
a variant of the original problem.
In this section we attempt to address this problem by defining a
decidable test on constraint problems and proving that this test
allows us to avoid reducing may-divergent problems.
We begin by introducing some terminology---we say that a
constraint problem $\cprob{\etenv}{\prob}$ is
\begin{iteMize}{$\bullet$}
  \item \emph{terminal} \pn{written $\cprob{\etenv}{\prob}\notrans$}
  if there does not exist a constraint problem
  $\cprob{\etenv'}{\prob'}$ such that
  $\cprob{\etenv}{\prob}\trans\cprob{\etenv'}{\prob'}$.
  \item a \emph{$\trans$-normal form} of $\cprob{\twinkle{\etenv}}{\twinkle{\prob}}$
  if there exists a finite transformation sequence from
  $\cprob{\twinkle{\etenv}}{\twinkle{\prob}}$ to $\cprob{\etenv}{\prob}$
  and $\cprob{\etenv}{\prob}$ is terminal.
  \item \emph{strongly normalising} if all transformation sequences
  starting from $\cprob{\etenv}{\prob}$ eventually reach a
  terminal constraint problem.
  \item \emph{may-divergent} if there exists an infinite
  transformation sequence starting from $\cprob{\etenv}{\prob}$.\smallskip
\end{iteMize}

Our termination check will involve translating
elements of $\Prob{\SIG}$ into a subset of
$\Prob{\SIG}$ which corresponds to first-order
unification problems.
Since first-order unification is known to be decidable,
we can check whether the first-order unification problem that
underlies a given non-permutative constraint problem is satisfiable.
From this we will deduce whether the non-permutative
constraint problem is strongly normalising.

We refer to this abstraction interpretation process
as \qtn{first-order reduction}, and begin by
reducing nominal signatures and types to first-order versions.

\begin{defi}[Reducing nominal signatures]
\label{def:fl-sig}
  For every nominal signature $\SIG$ we write $\fl{\SIG}$ for the
  underlying first-order signature, which has
  $\SIGNtys{\fl{\SIG}}\defeq\emptyset$ and
  $\SIGDtys{\fl{\SIG}}\defeq\SIGDtys{\SIG}$.
  Furthermore, if $\CONTYPE{\CON}{\ty}{\dty}{\SIG}$
  then $\CONTYPE{\CON}{\fl{\ty}}{\dty}{\fl{\SIG}}$ holds,
  where $\fl{\ty}$ is defined as follows.
  \begin{mathpar}
    \fl{\nty}\defeq\UNITTY \and
    \fl{\pn{\ABSTY{\nty}{\ty}}}\defeq\PRODTY{\UNITTY}{\pn{\fl{\ty}}} \and
    \fl{\dty}\defeq\dty \and
    \fl{\UNITTY}\defeq\UNITTY \and
    \fl{\pn{\TUPTY{\ty_1}{\ty_k}}}\defeq
    \TUPTY{\pn{\fl{\ty_1}}}{\pn{\fl{\ty_k}}}.
  \end{mathpar}\smallskip
\end{defi}

\noindent Note that we reduce all name sorts to the unit type
and all abstraction type-formers $\ABSTY{\nty}{\ty}$
to the product type $\PRODTY{\UNITTY}{\pn{\fl{\ty}}}$.
Thus we lose all information on the types of object-level names
but retain some information on the locations of abstractions
in the original type.

\begin{defi}[Reducing type environments]
\label{def:fl-etenv}
  For any type environment $\etenv$ we write $\fl{\etenv}$
  for the type environment such that
  \begin{iteMize}{$\bullet$}
    \item $\dom{\fl{\etenv}}=\setwhere{\var}{\text{$\var\in\dom{\etenv}$ and
    $\etenv\pn{\var}$ is \emph{not} a name sort}}$; and
    \item $\fl{\etenv}\pn{\var}=\fl{\pn{\etenv\pn{\var}}}$ for all
    $\var\in\dom{\fl{\etenv}}$.
  \end{iteMize}
\end{defi}

Note that $\dom{\fl{\etenv}}\subseteq\dom{\etenv}$ by definition.
We now define a similar first-order reduction operation
on non-permutative nominal terms---this definition
includes a type environment
$\etenv$ as a parameter, which is necessary to
decide how to handle variables when they are encountered
during the reduction process.
This definition is related to the morphism between nominal
and first-order terms defined in \cite{CalvesC:fironl}.

\begin{defi}[Reducing terms]
\label{def:fl-term}
  For a term $\term$, let $\etenv$ be any type environment
  such that $\expjb{\SIG}{\etenv}{\term}{\ty}$ holds for some $\ty$.
  Then, define the first-order reduction of $\term$ under $\etenv$,
  $\flb{\etenv}{\term}$, as follows.
  \begin{mathpar}
    \flb{\etenv}{\var}\defeq
    \begin{cases}
      \UNIT & \text{if $\etenv\pn{\var}$ is a name sort} \\
      \var  & \text{if $\etenv\pn{\var}$ is \emph{not} a name sort}
    \end{cases} \and
    \flb{\etenv}{\pn{\ABS{\var}{\term}}}\defeq\PAIR{\UNIT}{\flb{\etenv}{\term}} \and
    \flb{\etenv}{\UNIT}\defeq\UNIT \and
    \flb{\etenv}{\pn{\APP{\CON}{\term}}}\defeq
    \APP{\CON}{\pn{\flb{\etenv}{\term}}} \and
    \flb{\etenv}{\TUPLE{\term_1}{\term_k}}\defeq
    \TUPLE{\pn{\flb{\etenv}{\term_1}}}{\pn{\flb{\etenv}{\term_k}}}
  \end{mathpar}\smallskip
\end{defi}

\noindent Mirroring \Defn{\ref{def:fl-sig}}, we turn any variables of name sort
into unit terms and translate abstraction term-formers into a pair
consisting of unit and the reduced abstraction body.
This will be convenient later on because it ensures that the
\qtn{size} of a reduced term $\flb{\etenv}{\term}$ \pn{defined below}
is the same as the size of the original term $\term$.
It is straightforward to show that
if $\expjb{\SIG}{\etenv}{\term}{\ty}$ then
$\expjb{\fl{\SIG}}{\fl{\etenv}}{\flb{\etenv}{\term}}{\fl{\ty}}$.
We also get a \qtn{weakening} result:
if $\expjb{\SIG}{\etenv}{\term}{\ty}$
and $\etenv'\supseteq\etenv$ then
$\flb{\etenv'}{\term}=\flb{\etenv}{\term}$.
We now extend the definitions to constraint problems.

\begin{defi}[Reducing constraint problems]
  \label{def:fl-prob}
  The reduction $\flb{\etenv}{\prob}$ of constraints $\prob$
  under $\etenv$, where $\formjb{\SIG}{\etenv}{\prob}$, is defined as follows.
  \begin{eqnarray*}
    \flb{\etenv}{\prob} & \defeq &
    \setwhere{\ECON{\flb{\etenv}{\term_1}}{\flb{\etenv}{\term_2}}}
    {\pn{\ECON{\term_1}{\term_2}}\in\prob}.
  \end{eqnarray*}
  For a constraint problem $\cprob{\etenv}{\prob}$, we write
  $\fl{\pn{\cprob{\etenv}{\prob}}}$ for the corresponding reduced constraint problem
  $\cprob{\fl{\etenv}}{\flb{\etenv}{\prob}}$.
  It is trivial to show that if
  $\formjb{\SIG}{\emptyset}{\cprob{\etenv}{\prob}}$ holds then
  $\formjb{\fl{\SIG}}{\emptyset}{\fl{\pn{\cprob{\etenv}{\prob}}}}$
  also holds.
\end{defi}

When reducing constraint problems we discard any freshness
constraints, since these are not present in first-order
unification problems.
This is not an issue since freshness constraints cannot
cause may-divergence in our constraint transformation rules.
For equality constraints, we simply apply the
first-order reduction defined in \Defn{\ref{def:fl-term}}
to both terms separately.

In order to reason about the satisfaction of
reduced constraint problems we must define
first-order reductions of ground trees, \al-trees and
\al-tree valuations.

\begin{defi}[Reducing ground trees]
  \label{def:fl-tree}
  We define the first-order reduction $\fl{\tree}$ of a ground tree $\tree$
  as follows.
  \begin{mathpar}
    \fl{\name}\defeq\UNIT \and
    \fl{\UNIT}\defeq\UNIT \and
    \fl{\pn{\ABS{\name}{\tree}}}\defeq\PAIR{\UNIT}{\fl{\tree}} \and
    \fl{\pn{\APP{\CON}{\tree}}}\defeq\APP{\CON}{\pn{\fl{\tree}}} \and
    \fl{\TUPLE{\tree_1}{\tree_k}}\defeq
    \TUPLE{\fl{\tree_1}}{\fl{\tree_k}}.
  \end{mathpar}
\end{defi}\vspace{4 pt}

\noindent It is easy to show that $\tree\in\Tree{\SIG}{\ty}$
implies $\fl{\tree}\in\Tree{\fl{\SIG}}{\fl{\ty}}$
for any ground tree $\tree$.
Regarding \al-equivalence,
if $\tree\in\Tree{\SIG}{\ty}$ then
$\aec{\fl{\tree}}=\setb{\fl{\tree}}$ and
$\aec{\fl{\tree}}\in\Atree{\fl{\SIG}}{\fl{\ty}}$.
Furthermore, if $\aeq{\tree_1}{\tree_2}{\ty}$ then
$\fl{\tree_1}=\fl{\tree_2}$, since the first-order
reduction process erases all names.
This property means that we can define the
first-order reduction of \al-trees as follows:
$\fl{\aec{\tree}}=\aec{\fl{\tree}}=\setb{\fl{\tree}}$.
In turn, this allows us to define a first-order
reduction operation on \al-tree valuations.

\begin{defi}[Reducing \al-tree valuations]
  \label{def:fl-vln}
  If $\vln\in\Atree{\SIG}{\etenv}$ then we write
  $\fl{\vln}$ for the valuation which has
  $\dom{\fl{\vln}}=\dom{\fl{\etenv}}$
  \pn{and hence $\dom{\fl{\vln}}\subseteq\dom{\vln}$}
  and is such that $\fl{\vln}\pn{\var}=\fl{\pn{\vln\pn{\var}}}$
  for all $\var\in\dom{\fl{\vln}}$.
  It is straightforward to show that
  if $\vln\in\Atree{\SIG}{\etenv}$ then
  $\fl{\vln}\in\Atree{\fl{\SIG}}{\fl{\etenv}}$.
\end{defi}

We will use $\vlnw$ to range over valuations in $\Atree{\fl{\SIG}}{\fl{\etenv}}$
if the starting \al-tree valuation in $\Atree{\SIG}{\etenv}$ is irrelevant.
Now, a key task is to show that satisfaction is preserved by the process
of reduction to first-order form.

\begin{lem}[Reduction and valuation]
  \label{lem:fl-vapp}
  For any $\SIG$, $\vln$, $\etenv$, $\term$ and $\ty$,
  if $\expjb{\SIG}{\etenv}{\term}{\ty}$ and
  $\vln\in\Atree{\SIG}{\etenv}$ then
  $\fl{\pn{\vapp{\vln}{\term}}}=\vapp{\fl{\vln}}{\flb{\etenv}{\term}}$.
\end{lem}
\proof
  The proof is by induction on the structure of $\term$.
  In the base case where $\term$ is a variable which is not of name sort,
  we use the defining properties of $\fl{\vln}$.\qed

\begin{lem}[Reduction and satisfaction]
  \label{lem:fl-satis}
  For any $\SIG$, $\etenv$, $\prob$ and $\vln$, if
  $\formjb{\SIG}{\etenv}{\prob}$ and $\vln\in\Atree{\SIG}{\etenv}$ then
  $\SATIS{\vln}{\prob}$ implies $\SATIS{\fl{\vln}}{\flb{\etenv}{\prob}}$.
\end{lem}
\proof
  Since \Defn{\ref{def:fl-prob}} discards all freshness constraints
  in $\prob$ and translates all freshness constraints,
  it suffices to show that, for any equality constraint
  $\pn{\ECON{\term_1}{\term_2}}\in\prob$,
  if $\vapp{\vln}{\term_1}=\vapp{\vln}{\term_2}$ then
  $\vapp{\fl{\vln}}{\flb{\etenv}{\term_1}}=
   \vapp{\fl{\vln}}{\flb{\etenv}{\term_2}}$.
  If we assume that $\vapp{\vln}{\term_1}=\vapp{\vln}{\term_2}$ then
  $\fl{\pn{\vapp{\vln}{\term_1}}}=\fl{\pn{\vapp{\vln}{\term_2}}}$,
  and by \Lem{\ref{lem:fl-vapp}} we get that
  $\vapp{\fl{\vln}}{\flb{\etenv}{\term_1}}=
  \vapp{\fl{\vln}}{\flb{\etenv}{\term_2}}$,
  as required.\qed

We can now prove an important result about the satisfaction
of non-permutative constraint problems and state its corollary.

\begin{thm}[Reduction and satisfiability]
  \label{thm:fl-sat}
  For any $\SIG$, $\etenv$ and $\prob$,
  if $\formjb{\SIG}{\emptyset}{\cprob{\etenv}{\prob}}$
  then $\SAT{\cprob{\etenv}{\prob}}$ implies
  $\SAT{\fl{\pn{\cprob{\etenv}{\prob}}}}$.
\end{thm}
\proof
  If $\SAT{\cprob{\etenv}{\prob}}$ then there exists a valuation
  $\vln\in\Atree{\SIG}{\etenv}$ such that $\SATIS{\vln}{\prob}$.
  By \Lem{\ref{lem:fl-satis}} it follows that
  $\SATIS{\fl{\vln}}{\flb{\etenv}{\prob}}$, where
  $\fl{\vln}\in\Atree{\fl{\SIG}}{\fl{\etenv}}$.
  Thus we have $\SAT{\fl{\pn{\cprob{\etenv}{\prob}}}}$,
  as required.\qed

\begin{cor}[Reduction and unsatisfiability]
  \label{cor:fl-sat}
  For any $\SIG$, $\etenv$ and $\prob$,
  if $\formjb{\SIG}{\emptyset}{\cprob{\etenv}{\prob}}$
  then $\NOTSAT{\fl{\pn{\cprob{\etenv}{\prob}}}}$ implies
  $\NOTSAT{\cprob{\etenv}{\prob}}$.\qed
\end{cor}

\Cor{\ref{cor:fl-sat}} tells us that if the first-order reduction
$\fl{\pn{\cprob{\etenv}{\prob}}}$ is \emph{unsatisfiable}
then the original problem $\cprob{\etenv}{\prob}$ is unsatisfiable.
This is one of the properties that we require of a correct termination
check for non-permutative constraint problems.
It just remains to show that if the first-order reduction
$\fl{\pn{\cprob{\etenv}{\prob}}}$ is \emph{satisfiable} then
$\cprob{\etenv}{\prob}$ is strongly normalising.

We begin by showing that satisfaction of reduced
constraint problems is preserved by substitution,
in the following sense.

\begin{lem}[Satisfaction and substitution]
  \label{lem:fl-sub-sat}
  For any $\SIG$, $\etenv$, $\prob$, $\var$, $\term$ and $\vlnw$,
  suppose that $\formjb{\SIG}{\etenv}{\prob}$ and
  $\formjb{\SIG}{\etenv}{\ECON{\var}{\term}}$ and
  $\vlnw\in\Atree{\fl{\SIG}}{\fl{\etenv}}$.
  Then, if $\SATIS{\vlnw}{\flb{\etenv}{\prob}}$ and
  $\SATIS{\vlnw}{\ECON{\flb{\etenv}{\var}}{\flb{\etenv}{\term}}}$
  then $\SATIS{\vlnw}{\flb{\etenv}{\pn{\sapp{\sub{\var}{\term}}{\prob}}}}$.
\end{lem}
\proof
  If $\SATIS{\vlnw}{\ECON{\flb{\etenv}{\var}}{\flb{\etenv}{\term}}}$ then we
  know that $\vapp{\vlnw}{\flb{\etenv}{\var}}=\vapp{\vlnw}{\flb{\etenv}{\term}}$.
  By \Defn{\ref{def:fl-prob}} it suffices to prove that a similar
  substitution property holds for all equality constraints
  $\pn{\ECON{\term_1}{\term_2}}\in\prob$.
  We assume that $\expjb{\SIG}{\etenv}{\term_1}{\ty}$ and
  $\expjb{\SIG}{\etenv}{\term_2}{\ty}$ \pn{for some $\ty$} and that
  $\formjb{\SIG}{\etenv}{\ECON{\var}{\term}}$ and
  $\vlnw\in\Atree{\fl{\SIG}}{\fl{\etenv}}$, and show that
  $\vapp{\vlnw}{\flb{\etenv}{\term_1}}=\vapp{\vlnw}{\flb{\etenv}{\term_2}}$ and
  $\vapp{\vlnw}{\flb{\etenv}{\var}}=\vapp{\vlnw}{\flb{\etenv}{\term}}$ imply
  $\vapp{\vlnw}{\flb{\etenv}{\pn{\sapp{\sub{\var}{\term}}{\term_1}}}}=
   \vapp{\vlnw}{\flb{\etenv}{\pn{\sapp{\sub{\var}{\term}}{\term_2}}}}$.
  We now perform a case split: if $\ty$ is a name sort,
  the result follows from the fact that $\flb{\etenv}{\term'}=
  \flb{\etenv}{\pn{\sapp{\sub{\var}{\term}}{\term'}}}$
  if $\term$ and $\var$ are both of name sort;
  otherwise it follows from \Lem{\ref{lem:sub-vln}}
  and the fact that
  $\flb{\etenv}{\pn{\sapp{\sub{\var}{\term}}{\term'}}}=
  \sapp{\sub{\var}{\flb{\etenv}{\term}}}{\flb{\etenv}{\term'}}$.\qed

We can now show that solutions to reduced problems are
preserved by reduction of the original problem.

\begin{lem}[Preservation of reduced solutions]
  \label{lem:fl-trans}
  If $\formjb{\SIG}{\emptyset}{\cprob{\etenv}{\prob}}$ and
  $\vlnw\in\Atree{\fl{\SIG}}{\fl{\etenv}}$ and
  $\SATIS{\vlnw}{\flb{\etenv}{\prob}}$ and
  $\cprob{\etenv}{\prob}\trans\cprob{\etenv'}{\prob'}$
  then there exists a valuation
  $\vlnw'\in\Atree{\fl{\SIG}}{\fl{\etenv'}}$
  which agrees with $\vlnw$ on $\dom{\fl{\etenv}}$
  and is such that $\SATIS{\vlnw'}{\flb{\etenv'}{\prob'}}$.
\end{lem}
\proof
  We assume that $\formjb{\SIG}{\emptyset}{\cprob{\etenv}{\prob}}$ and
  $\vlnw\in\Atree{\fl{\SIG}}{\fl{\etenv}}$ and
  $\SATIS{\vlnw}{\flb{\etenv}{\prob}}$ and
  $\cprob{\etenv}{\prob}\trans\cprob{\etenv'}{\prob'}$ all hold,
  and proceed by case analysis on the constraint transformation rule
  used to derive $\cprob{\etenv}{\prob}\trans\cprob{\etenv'}{\prob'}$.

  The cases for rules \frule{1}--\frule{3} and \frule{5} are
  straightforward since $\flb{\etenv'}{\prob'}=\flb{\etenv}{\prob}$.
  In the case for rule \erule{1} the new constraints in $\flb{\etenv'}{\prob'}$
  are all trivially satisfied. 
  The cases for rules \erule{2} and \erule{3} follow directly from the
  semantics of non-permutative constraints, defined in terms of $\vapp{\vln}{\term}$.
  The case for \erule{5} follows because $\flb{\etenv'}{\prob'}$ is obtained from
  $\flb{\etenv}{\prob}$ simply by deleting a constraint.
  In the cases for \frule{4} and \erule{4} the additional constraints in
  $\flb{\etenv'}{\prob'}$ are all simple equality constraints in involving
  tuples and unit, which are trivially satisfied.
  The cases for \erule{6} and \erule{7} both rely on
  \Lem{\ref{lem:fl-sub-sat}} to deal with substitution.
  In the case of \erule{7} we also use the fact that the
  variables in the \qtn{patterns} generated by the narrowing
  rules from \Fig{\ref{fig:narrow}} always use fresh variables.
  This allows us to safely extend $\vlnw$ to produce a larger
  valuation $\vlnw'$.\qed

In order to prove a termination result we will need to define
some kind of size metric on constraint problems, by interpreting
them into a well-founded set.

\begin{defi}[Sizes of ground trees]
  \label{def:fl-size-tree}
  Let $\NatPlus$ be the set $\setwhere{n\in\Nat}{n\geq 1}$.
  Then we define a size function $\size{\tree}$ which maps
  from ground trees into $\NatPlus$, as follows.
  \begin{mathpar}
    \size{\name}\defeq1 \and
    \size{\APP{\CON}{\tree}}\defeq1+\size{\tree} \and
    \size{\ABS{\var}{\tree}}\defeq2+\size{\tree} \\
    \size{\UNIT}\defeq1 \and
    \size{\TUPLE{\tree_1}{\tree_k}}\defeq
    1+\bigsum{i\in\setrange{1}{k}}{\size{\tree_i}}.
  \end{mathpar}
  Since $\aeqnt{\tree}{\tree'}$ implies that
  $\size{\tree}=\size{\tree'}$, the above definition
  induces a well-defined size function on
  \al-trees: $\size{\aec{\tree}}=\size{\tree}$
  for some/any representative $\tree$ of the
  \al-equivalence class.
  Furthermore, $\size{\tree}\geq 1$ for all $\tree$.
\end{defi}

\begin{defi}[Sizes of terms and atomic constraints]
  \label{def:fl-size-term}
  If $\expjb{\SIG}{\etenv}{\term}{\ty}$ \pn{for some type $\ty$}
  then we define a function $\termeq{\term}\OFTY
  \FUNTY{\pn{\Atree{\fl{\SIG}}{\fl{\etenv}}}}{\NatPlus}$ as follows.
  \begin{mathpar}
    \termsize{\var}{\vlnw}\defeq
    \begin{cases}
      1 & \text{if $\etenv\pn{\var}=\nty$, for some $\nty$} \\
      \size{\vlnw\pn{\var}} & \text{otherwise}
    \end{cases} \and
    \termsize{\ABS{\var}{\term}}{\vlnw}\defeq
    2 + \termsize{\term}{\vlnw} \\
    \termsize{\UNIT}{\vlnw}\defeq 1 \and
    \termsize{\APP{\CON}{\term}}\defeq
    1 + \termsize{\term}{\vlnw} \and
    \termsize{\TUPLE{\term_1}{\term_k}}{\vlnw}\defeq
    1 + \bigsum{i\in\setrange{1}{k}}{\termsize{\term_i}{\vlnw}}.
  \end{mathpar}
  We now extend this definition to atomic constraints---if
  $\formjb{\SIG}{\etenv}{\constr}$ we define a function
  $\termeq{\constr}\OFTY\FUNTY{\pn{\Atree{\fl{\SIG}}{\fl{\etenv}}}}{\NatPlus}$,
  as follows.
  \begin{mathpar}
    \termsize{\ECON{\term}{\term'}}{\vlnw} \defeq
    \termsize{\term}{\vlnw} + \termsize{\term'}{\vlnw} \and
    \termsize{\FCON{\var}{\term'}}{\vlnw} \defeq \termsize{\term'}{\vlnw}.
  \end{mathpar}
\end{defi}

Note that $\termsize{\term}{\vlnw}\geq 1$ and $\termsize{\constr}{\vlnw}\geq 1$
both always hold.

\begin{lem}[Equality constraints and sizes]
  \label{lem:econ-sizes}
  For all $\SIG$, $\etenv$, $\term$, $\term'$ and $\vlnw$,
  if $\formjb{\SIG}{\etenv}{\ECON{\term}{\term'}}$
  and $\vlnw\in\Atree{\fl{\SIG}}{\fl{\etenv}}$
  and $\SATIS{\vlnw}{\ECON{\flb{\etenv}{\term}}{\flb{\etenv}{\term'}}}$
  then $\termsize{\term}{\vlnw}=\termsize{\term'}{\vlnw}$.
\end{lem}
\proof
  If $\SATIS{\vlnw}{\ECON{\flb{\etenv}{\term}}{\flb{\etenv}{\term'}}}$
  then $\vapp{\vlnw}{\flb{\etenv}{\term}}=\vapp{\vlnw}{\flb{\etenv}{\term'}}$.
  Hence $\size{\vapp{\vlnw}{\flb{\etenv}{\term}}}=
  \size{\vapp{\vlnw}{\flb{\etenv}{\term'}}}$.
  It is straightforward to show that
  $\size{\vapp{\vlnw}{\flb{\etenv}{\term}}}=\termsize{\term}{\vlnw}$
  \pn{the crucial cases are for names and abstractions} and hence
  we get that $\termsize{\term}{\vlnw}=\termsize{\term'}{\vlnw}$,
  as required.\qed

\begin{defi}[Solved variables]
  \label{def:constraint-solved-var}
  We say that a variable $\var$ is \emph{solved} in
  $\cprob{\etenv}{\prob}$ iff there is precisely one
  occurrence of $\var$ in $\prob$, where that occurrence is
  in a constraint of the form $\ECON{\var}{\term}$ or $\ECON{\term}{\var}$.
  We say that a variable is \emph{unsolved} when it is not solved.
\end{defi}

\begin{defi}[Measure on constraint problems]
  \label{def:fl-meas}
  Write $\NatMS$ for the set of finite multisets of elements
  of $\NatPlus$ and write $\msetwhere{f\pn{x}}{x\in S, P\pn{x}}$
  for the \emph{multiset} of values $f\pn{x}$ where $x\in S$ and
  $x$ satisfies the property $P\pn{x}$.
  Now, we begin by defining two intermediate measure functions
  from constraint problems $\cprob{\etenv}{\prob}$ into $\NatMS$,
  each parameterised by a valuation $\vlnw\in\Atree{\fl{\SIG}}{\fl{\etenv}}$.
  \begin{eqnarray*}
    \measa{\vlnw}{\cprob{\etenv}{\prob}} & \defeq &
    \msetwhere{\termsize{\var}{\vlnw}}
    {\var\in\dom{\etenv}, \text{$\var$ is \emph{unsolved}}} \\
    \measb{\vlnw}{\cprob{\etenv}{\prob}} & \defeq &
    \msetwhere{\termsize{\constr}{\vlnw}}{\constr\in\prob}
  \end{eqnarray*}
  We now define a measure function $\meassym$ on constraint problems
  in terms of $\meassym_1$, $\meassym_2$ and a valuation
  $\vlnw\in\Atree{\fl{\SIG}}{\fl{\etenv}}$.
  \begin{eqnarray*}
    \meas{\vlnw}{\cprob{\etenv}{\prob}} & \defeq &
    \pair{\measa{\vlnw}{\cprob{\etenv}{\prob}}}
    {\measb{\vlnw}{\cprob{\etenv}{\prob}}}
  \end{eqnarray*}
  Since $\meassym_1$ and $\meassym_2$ are functions into $\NatMS$
  it follows that $\meassym$ is a function into $\cartesian{\NatMS}{\NatMS}$.
  There is a well-founded ordering $\mordsubsym{\NatMS}$ on $\NatMS$
  induced by the usual well-founded ordering $<$ on natural numbers,
  via the multiset ordering construction from \cite{DershowitzN:protwm}.
  From the lexicographic product of $\mordsubsym{\NatMS}$ with itself
  we derive a well-founded ordering $\mordsubsym{\cartesian{\NatMS}{\NatMS}}$
  on $\cartesian{\NatMS}{\NatMS}$, which we will use to provide a
  well-founded ordering on the results of the $\meassym$ function.
\end{defi}

We now have the necessary tools to show that transformation of
any constraint problem whose first-order reduction is satisfiable
will terminate.
The proof uses the standard strategy of interpreting
constraint problems in a set equipped with a well-founded ordering,
using the measure function defined above.

\begin{thm}[Termination]
\label{thm:constraint-term}
  If $\formjb{\SIG}{\emptyset}{\cprob{\etenv}{\prob}}$ holds and
  $\SAT{\fl{\pn{\cprob{\etenv}{\prob}}}}$ then there is
  no infinite sequence of $\trans$ transformations starting from
  $\cprob{\etenv}{\prob}$.
\end{thm}
\proof
  We will proceed by showing that
  \begin{equation}
    \label{eq:constraint-term-tgt}
    \cprob{\etenv}{\prob}\trans\cprob{\etenv'}{\prob'} \;\imp\;
    \mordsub{\cartesian{\NatMS}{\NatMS}}{\meas{\vlnw'}{\cprob{\etenv'}{\prob'}}}
    {\meas{\vlnw}{\cprob{\etenv}{\prob}}}
  \end{equation}
  holds, for any $\vlnw\in\Atree{\fl{\SIG}}{\fl{\etenv}}$ such that
  $\SATIS{\vlnw}{\flb{\etenv}{\prob}}$ and any $\vlnw'\in\Atree{\fl{\SIG}}{\fl{\etenv'}}$
  which agrees with $\vlnw$ on $\dom{\fl{\etenv}}$ and is such that
  $\SATIS{\vlnw'}{\flb{\etenv'}{\prob'}}$.

  This suffices to prove termination because \pn{by \Lem{\ref{lem:fl-trans}}}
  solutions to reduced problems are preserved by the transformation rules.
  By assumption, if $\SAT{\fl{\pn{\cprob{\etenv}{\prob}}}}$ holds then
  there exists some $\vlnw\in\Atree{\fl{\SIG}}{\fl{\etenv}}$
  such that $\SATIS{\vlnw}{\flb{\etenv}{\prob}}$.
  Therefore, for any solution $\vlnw$ of the reduced problem
  $\fl{\pn{\cprob{\etenv}{\prob}}}$ we get a family of derived
  solutions $\vlnw'$ for the reduced problem
  $\fl{\pn{\cprob{\etenv'}{\prob'}}}$, each of which produces a
  strictly smaller value for $\meas{\vlnw'}{\cprob{\etenv'}{\prob'}}$
  in the well-founded ordering $\mordsubsym{\cartesian{\NatMS}{\NatMS}}$.
  If we repeat this argument along the transformation
  sequence it follows that the chain must eventually terminate.

  To prove that \eqref{eq:constraint-term-tgt} holds,
  we proceed by case analysis on the rule used to derive
  $\cprob{\etenv}{\prob}\trans\cprob{\etenv'}{\prob'}$.
  We will present the case for rule \erule{7} in full,
  as it is the most involved.
  \begin{desCription}

    \item\noindent{\hskip-12 pt\bf \erule{7}:}\
      We assume that $\prob=\CONJ{\pn{\ECON{\abss{\var}{1}{k}{\var}}
      {\abss{\vy}{1}{k}{\term}}}}{\twinkle{\prob}}$
      and that $\SATIS{\vlnw}{\flb{\etenv}{\prob}}$,
      where $\flb{\etenv}{\prob}=\CONJ{\flb{\etenv}
      {\pn{\ECON{\abss{\var}{1}{k}{\var}}{\abss{\vy}{1}{k}{\term}}}}}
      {\flb{\etenv}{\twinkle{\prob}}}$ and
      $\vlnw\in\Atree{\fl{\SIG}}{\fl{\etenv}}$.
      We also assume that
      $\prob'=\CONJ{\pn{\ECON{\var}{\twinkle{\term}}}}{\CONJ
      {\pn{\ECON{\abss{\var}{1}{k}{\var}}{\abss{\vy}{1}{k}{\term}}}}
      {\pn{\sapp{\sub{\var}{\twinkle{\term}}}{\twinkle{\prob}}}}}$
      and $\etenv'=\etenv,\twinkle{\etenv}$ both hold,
      where $\narrow{\etenv}{\term}{\twinkle{\etenv}}{\twinkle{\term}}$ holds.
      Furthermore, we let $\vlnw'\in\Atree{\fl{\SIG}}{\fl{\etenv'}}$
      be such that $\SATIS{\vlnw'}{\flb{\etenv'}{\prob'}}$ holds
      and such that $\vlnw$ and $\vlnw'$ agree on $\dom{\fl{\etenv}}$.
      By assumption, $\var\notin\vars{\term}$ and since
      $\dom{\twinkle{\etenv}}\inter\dom{\etenv}=\emptyset$ by definition,
      it follows that rule \erule{7} changes $\var$ from unsolved to solved
      and replaced it with unsolved variables $\dom{\twinkle{\etenv}}$.
      Thus we must show that $\termsize{\vz}{\vlnw'}<\termsize{\var}{\vlnw}$
      for all $\vz\in\dom{\twinkle{\etenv}}$.

      Now, since $\SATIS{\vlnw'}{\flb{\etenv'}{\prob'}}$ we know that
      $\SATIS{\vlnw'}{\ECON{\flb{\etenv'}{\var}}{\flb{\etenv'}{\twinkle{\term}}}}$ holds,
      and by \Lem{\ref{lem:econ-sizes}} we get that
      $\termsize{\var}{\vlnw'}=\termsize{\twinkle{\term}}{\vlnw'}$ holds.
      Since $\term$ is not a variable we know that $\etenv\pn{\var}$
      cannot be a name sort and hence $\var\in\dom{\fl{\etenv}}$.
      Thus we get that $\vlnw\pn{\var}=\vlnw'\pn{\var}$
      and it follows that $\termsize{\var}{\vlnw}=\termsize{\var}{\vlnw'}$.
      Therefore we know that $\termsize{\var}{\vlnw}=\termsize{\twinkle{\term}}{\vlnw'}$.
      It is easy to see that $\termsize{\vz}{\vlnw'}<\termsize{\twinkle{\term}}{\vlnw'}$
      holds for all $\vz\in\dom{\twinkle{\etenv}}$, from which it follows that
      $\termsize{\vz}{\vlnw'}<\termsize{\var}{\vlnw}$
      holds for all $\vz\in\dom{\twinkle{\etenv}}$.
      Hence we have shown that
      $\mordsub{\NatMS}{\measa{\vlnw'}{\cprob{\etenv'}{\prob'}}}
      {\measa{\vlnw}{\cprob{\etenv}{\prob}}}$ holds,
      from which it follows that $\mordsub{\cartesian{\NatMS}{\NatMS}}
      {\meas{\vlnw'}{\cprob{\etenv'}{\prob'}}}
      {\meas{\vlnw}{\cprob{\etenv}{\prob}}}$, as required.\smallskip

  \end{desCription}

  \noindent In the case for rule \erule{6}, one unsolved variable
  becomes solved.
  In each of the remaining cases, there are no more unsolved variables in
  $\dom{\etenv'}$ than in $\dom{\etenv}$.
  Furthermore, $\measb{\vlnw'}{\cprob{\etenv'}{\prob'}}$ is formed from
  $\measb{\vlnw}{\cprob{\etenv}{\prob}}$ by removing
  $\termsize{\twinkle{\constr}}{\vlnw}$ \pn{for some $\twinkle{\constr}$}
  and replacing it with zero or finitely many elements
  $\termsize{\twinkle{\constr}_i}{\vlnw'}$, where $i\in\setrange{1}{k}$ for some $k$,
  and where $\vlnw'=\vlnw$ since $\etenv'=\etenv$.
  Now we get that $\termsize{\twinkle{\constr}_i}{\vlnw'}<
  \termsize{\twinkle{\constr}}{\vlnw}$ for all $i\in\setrange{1}{k}$.
  Thus we get
  $\mordsub{\NatMS}{\measb{\vlnw'}{\cprob{\etenv'}{\prob'}}}
  {\measb{\vlnw}{\cprob{\etenv}{\prob}}}$, and it follows that
  $\mordsub{\cartesian{\NatMS}{\NatMS}}
  {\meas{\vlnw'}{\cprob{\etenv'}{\prob'}}}
  {\meas{\vlnw}{\cprob{\etenv}{\prob}}}$, as required.\qed

Thus we have shown that \textit{\pn{1}}
if the reduced problem $\fl{\pn{\cprob{\etenv}{\prob}}}$
is \emph{satisfiable} then the original problem
$\cprob{\etenv}{\prob}$ is strongly normalising, and \textit{\pn{2}}
if $\fl{\pn{\cprob{\etenv}{\prob}}}$ is
\emph{unsatisfiable} then $\cprob{\etenv}{\prob}$ is unsatisfiable.
These properties mean that satisfiability of reduced problems is a suitable
decidable check for detecting may-divergent constraint problems.

\subsection{Soundness and completeness of the algorithm}
\label{subsec:correctness-algorithm}

In this section we prove correctness of a decision procedure
for non-permutative nominal constraint problems
which uses the constraint transformation rules from \Fig{\ref{fig:trans}}.
We proceed by relating the syntactic forms of constraint problems
in $\trans$-normal form to their satisfiability.

\begin{defi}[Solved constraint problems]
\label{def:constraint-solved-prob}
  A constraint problem $\cprob{\etenv}{\prob}$ is \emph{solved}
  iff all constraints in $\prob$ have one of the following forms.
  \begin{enumerate}[(1)]
    \item\label{item:solved-1}
    $\FCON{\var}{\vy}$, where $\var$ and $\vy$ are distinct variables
    and either $\etenv\pn{\var}=\etenv\pn{\vy}$ or
    $\etenv\pn{\vy}$ is not a name sort;
    \item\label{item:solved-2}
    $\ECON{\var}{\term}$, where $\var\notin\vars{\term}$ and
    $\var$ does not appear elsewhere in $\prob$;
    \item\label{item:solved-3}
    $\ECON{\abss{\var}{1}{k}{\var}}{\abss{\vy}{1}{k}{\vy}}$, where $k>0$ and
    $\etenv\pn{\var}\pn{=\etenv\pn{\vy}}$ is not a name sort
    and $\var$ and $\vy$ are distinct variables; or
    \item\label{item:solved-4}
    $\ECON{\abss{\var}{1}{k}{\var}}{\abss{\vy}{1}{k}{\var}}$, where $k>0$ and
    $\etenv\pn{\var}$ is not a name sort.
  \end{enumerate}
\end{defi}

\begin{lem}
\label{lem:constraint-solved-terminal}
  Any \emph{solved} constraint problem $\cprob{\etenv}{\prob}$
  is also \emph{terminal}.
\end{lem}
\proof
  By cases on the possible constraints that may appear
  within a solved constraint problem, according to
  \Defn{\ref{def:constraint-solved-prob}}.\qed

The relationship between terminal and solved constraints
and their satisfiability is crucial to the correctness of
our algorithm.
We now show that once a problem has been reduced as far as
possible using $\trans$ we can determine whether it is
satisfiable by examining its syntax.

\begin{lem}[Terminal constraints and satisfiability]
\label{lem:constraint-terminalsat}
  Let $\cprob{\etenv}{\prob}$ be a \emph{terminal} constraint problem
  such that $\formjb{\SIG}{\emptyset}{\cprob{\etenv}{\prob}}$.
  Then $\cprob{\etenv}{\prob}$ is satisfiable iff it is \emph{solved}.
\end{lem}
\proof
  We assume that $\formjb{\SIG}{\emptyset}{\cprob{\etenv}{\prob}}$.
  By inspection of the constraint transformation rules,
  the possible forms of constraint in a terminal constraint problem
  consist of the possibilities presented in
  \Defn{\ref{def:constraint-solved-prob}} as well as the following:
  \begin{enumerate}[(1)]
    \setcounter{enumi}{4}
    \item\label{item:terminal-5}
    $\FCON{\var}{\var}$.
    \item\label{item:terminal-6}
    $\ECON{\abss{\var}{1}{k}{\APP{\CON}{\term}}}
    {\abss{\vy}{1}{k}{\APP{\CON'}{\term'}}}$, where $\CON\neq\CON'$.
    \item\label{item:terminal-7}
    $\ECON{\var}{\term}$, where $\var\in\vars{\term}$ and $\term$ is not $\var$.
    \item\label{item:terminal-8}
    $\ECON{\abss{\var}{1}{k}{\var}}{\abss{\vy}{1}{k}{\term}}$,
    where $k>0$ and $\var\in\vars{\term}$.
  \end{enumerate}
  In particular, an equality constraint between two terms
  which have different numbers of outermost nested abstractions
  is not terminal, as it can be reduced by narrowing
  using rule \erule{7}.
  It suffices to show that any single constraint
  conforming to possibilities
  \ref{item:terminal-5}--\ref{item:terminal-8} is unsatisfiable,
  and that any solved constraint problem is satisfiable.
  We prove these below.
  \begin{desCription}

    \item\noindent{\hskip-12 pt\bf Any constraint of the form
      \ref{item:terminal-5}--\ref{item:terminal-8} is unsatisfiable:}\
      Constraints of form \ref{item:terminal-5} are not satisfiable
      because a name cannot be fresh for itself, and
      constraints of form \ref{item:terminal-6} are not satisfiable
      because the constructors do not match.
      Finally, constraints of forms \ref{item:terminal-7} and
      \ref{item:terminal-8} are not satisfiable because the
      occurs check fails.

    \item\noindent{\hskip-12 pt\bf Any solved constraint problem is
      satisfiable:}\ 
      For a solved constraint problem $\cprob{\etenv}{\prob}$ we will
      construct a satisfying valuation $\vln$.
      We write $\prob_i$ for the partition of $\prob$ where the
      constraints are all of the form $i\in\setb{1,2,3,4}$.

      We observe that we can form a satisfying valuation
      $\vln_{3,4}$ for $\prob_3\disu\prob_4$ because the
      variables $\var$ and $\vy$ in constraints of form 3
      and $\var$ in form 4 cannot be of name sort and hence
      cannot coincide with any of the abstracted variables.
      Therefore we can simply instantiate the abstracted
      variables distinctly \pn{i.e. avoiding aliasing}
      and instantiate the variables within the nesting
      to avoid the abstracted variables and satisfy the
      appropriate constraints.

      Now we note that if $\pn{\FCON{\var}{\vy}}\in\prob_1$
      and $\var,\vy\in\vars{\prob_3\disu\prob_4}$ then
      $\SATIS{\vln_{3,4}}{\FCON{\var}{\vy}}$ by construction.
      Therefore we can extend $\vln_{3,4}$ with additional mappings
      which ensure that all freshnesss in $\prob_1$ are satisfied,
      to produce a valuation $\vln_{1,3,4}$ which satisfies
      $\prob_1\disu\prob_3\disu\prob_4$.

      Finally it is always possible to extend $\vln_{1,3,4}$
      to a satisfying valuation $\vln$ for the entire problem $\prob$.
      We begin by providing an arbitrary instantiation for any variable
      $\vz\in\vars{\prob_2}$ which only appears on the right-hand side
      of constraints in $\prob_2$ and which has not already been instantiated.
      This just leaves the variables $\var$ which appear on the
      left-hand side of the constraints in $\prob_2$.
      By assumption on solved constraints these variables cannot
      appear elsewhere in $\prob$ and hence cannot have been
      instantiated already.
      Hence we are free to choose instantiations for these variables
      which satisfy $\prob_2$.
      Thus we get that $\SATIS{\vln}{\prob}$, as required.\qed

  \end{desCription}

\noindent With these results under our belt we can begin to examine the correctness
of the constraint transformation algorithm.
We begin by proving a soundness result: if a constraint problem
has a solved $\trans$-normal form then it is satisfiable.

\begin{thm}[Soundness of transformation algorithm]
\label{thm:alg-snd}
  For any constraint problem $\cprob{\etenv}{\prob}$
  where $\formjb{\SIG}{\emptyset}{\cprob{\etenv}{\prob}}$ holds,
  if there exists a $\trans$-normal form $\cprob{\etenv'}{\prob'}$
  of $\cprob{\etenv}{\prob}$ which is \emph{solved} then
  $\cprob{\etenv}{\prob}$ is satisfiable.
\end{thm}
\proof
  Suppose that $\cprob{\etenv'}{\prob'}$ is a $\trans$-normal
  form of $\cprob{\etenv}{\prob}$.
  If $\cprob{\etenv'}{\prob'}$ is solved then
  $\cprob{\etenv'}{\prob'}$ is satisfiable
  by \Lem{\ref{lem:constraint-terminalsat}},
  i.e.~there exists some $\vln'\in\Atree{\SIG}{\etenv'}$
  such that $\SATIS{\vln'}{\prob'}$.
  Finally, by \Thm{\ref{thm:constraint-sound}} we get that
  $\SATIS{\vln}{\prob}$ holds \pn{where $\vln$ is the
  restriction of $\vln'$ to $\dom{\etenv}$} and hence that
  $\cprob{\etenv}{\prob}$ is satisfiable, as required.\qed

We now prove a partial completeness result which is \emph{not quite}
the converse of \Thm{\ref{thm:alg-snd}} because it only applies to
strongly normalising constraint problems.
The assumption that $\cprob{\etenv}{\prob}$ is strongly normalising
is needed to ensure that it has a $\trans$-normal form,
which we then show is satisfied by $\vln$.

\begin{thm}[Partial completeness of transformation algorithm]
\label{thm:alg-cmp-term}
  Let $\cprob{\etenv}{\prob}$ be a strongly normalising
  constraint problem such that
  $\formjb{\SIG}{\emptyset}{\cprob{\etenv}{\prob}}$ holds.
  If $\cprob{\etenv}{\prob}$ is satisfiable then there
  exists a $\trans$-normal form $\cprob{\etenv'}{\prob'}$ of
  $\cprob{\etenv}{\prob}$ which is \emph{solved}.
\end{thm}
\proof
  If $\cprob{\etenv}{\prob}$ is satisfiable then there exists a valuation
  $\vln\in\Atree{\SIG}{\etenv}$ such that $\SATIS{\vln}{\prob}$ holds.
  Since $\cprob{\etenv}{\prob}$ is strongly normalising
  we know that every transformation sequence eventually terminates.
  Then, by \Thm{\ref{thm:constraint-complete}} we know that there is some
  sequence of transformations from $\cprob{\etenv}{\prob}$ which terminate
  at a problem $\cprob{\etenv'}{\prob'}$ such that $\SATIS{\vln'}{\prob'}$
  holds, where $\vln'$ extends $\vln$ to $\dom{\etenv'}$.
  Finally, by \Lem{\ref{lem:constraint-terminalsat}} it follows that
  the $\trans$-normal form $\cprob{\etenv'}{\prob'}$ is solved,
  as required.\qed

Now we use the termination checking procedure from
\Sect{\ref{subsec:correctness-termination}} to close the gap
in \Thm{\ref{thm:alg-cmp-term}}, giving us a correct decision
procedure for \NonPermSat{}.

\begin{thm}[Correct decision procedure]
\label{thm:alg-decid}
  There exists a correct decision procedure for \NonPermSat{}
  based on the constraint transformation rules from \Fig{\ref{fig:trans}}.
\end{thm}
\proof
  Using the termination check from \Thm{\ref{thm:constraint-term}}
  we can dismiss may-divergent constraint problems as unsatisfiable
  without having to rewrite them using the rules from \Fig{\ref{fig:trans}}.
  This allows us to restrict our attention to strongly normalising
  constraint problems $\cprob{\etenv}{\prob}$, for which we can compute
  the finite set $\sset$ of $\trans$-normal forms, that is, the set
  $\sset=\setwhere{\cprob{\etenv'}{\prob'}}
  {\cprob{\etenv}{\prob}\trans\cdots\trans\cprob{\etenv'}{\prob'}\notrans}$,
  in finite time.
  By \Thm{\ref{thm:alg-snd}} and \Thm{\ref{thm:alg-cmp-term}},
  the constraint problem $\cprob{\etenv}{\prob}$ is satisfiable
  precisely when there exists a solved constraint problem in $\sset$,
  which is a decidable property of the syntax of $\sset$.\qed

The algorithm presented in this section decides satisfiability of
non-permutative nominal constraint problems: it does not enumerate solutions.
Recalling \Defn{\ref{def:constraint-solved-prob}} we see that
solved constraint problems may contain constraints of the form
$\ECON{\abss{\var}{1}{k}{\var}}{\abss{\vy}{1}{k}{\vy}}$ or
$\ECON{\abss{\var}{1}{k}{\var}}{\abss{\vy}{1}{k}{\var}}$,
where $k>0$ and where $\var$ and $\vy$ are not variables of name sort.
Constraints of these forms may be satisfied by infinitely
many different ground valuations, as $\var$ and $\vy$ may range over some
recursive datatype.
Since these may be the only occurrences of $\var$ and $\vy$
in the constraint problem, it follows that a satisfiable
non-permutative nominal constraint problem may have
infinitely many satisfying ground valuations.
However, \Thm{\ref{thm:alg-decid}} demonstrates that our algorithm
only needs to check the finite number of elements of $\sset$ to ascertain
the patterns that all satisfying ground valuations must follow.

\section{Encoding name-name equivariant unification}
\label{sec:equivariant}

In this section we present a reduction of equivariant unification
between name-terms into non-permutative nominal constraints.
This will make explicit the link between equivariant unification
and non-permutative nominal constraints alluded to in
\Sect{\ref{sec:syntax-semantics}}.
It is sufficient for our purposes to consider
equivariant unification between terms of name sort because
that sub-problem of equivariant unification is known
to be \NP{}-complete \cite[\Sect{7.1}]{CheneyJ:nomlp-phd}.
We recall the grammar of equivariant unification name-terms
from \Sect{\ref{subsec:background-eu}}.
\begin{displaymath}
  \begin{array}{r@{\quad}rcl@{\qquad}l}
    \text{Vertices} & \vv,\vw & \bnf
            & \name & \text{\pn{name}} \\
    && \mid & \vara & \text{\pn{name variable}} \\[\jot]
    \text{Name-terms} & \va,\vb & \bnf
            & \pact{\PERM}{\vv} & \text{\pn{suspended permutation}} \\[\jot]
    \text{Permutation-terms} & \PERM & \bnf
            & \idp & \text{\pn{identity}} \\
    && \mid & \swap{\va}{\vb} & \text{\pn{swap}} \\
    && \mid & \pvar & \text{\pn{permutation variable}}
  \end{array}
\end{displaymath}
We also refer back to \Sect{\ref{subsec:background-eu}} for
the semantics of name-name equivariant unification problems.
To simplify our presentation we assume \pn{without loss of generality}
that all sub-terms of the forms $\pact{\pinv{\PERM}}{\vv}$ and
$\pact{\pn{\pcomp{\PERM}{\PERM'}}}{\vv}$
have been expanded away by the addition of fresh name variables
\pn{to represent intermediate values} and additional equality constraints.
This process is described as \qtn{phase two} of the equivariant
unification algorithm \cite[\Sect{4.2.2}]{CheneyJ:equu-jv}.
For example, the name-term $\pact{\pinv{\PERM}}{\vv}$ can be translated to the
fresh name variable $\vara$, given the constraint that
$\ECON{\pact{\PERM}{\vara}}{\vv}$ \pn{where $\PERM$ has been recursively
expanded out in the same way}.
Furthermore, we assume that all names are of a single name sort $\nty$.

\subsection{Defining the encoding}
\label{subsec:eu-define}

In order to encode equivariant unification we must use different collections
of variables to represent names, name variables and
permutation variables.
Thus we assume that the countably infinite set of variables $\Var$ is partitioned
into finitely many disjoint, countably infinite subsets $\VarName$, $\VarNvar$, and
$\lrange{\VarPvar{\pvar_1}}{\VarPvar{\pvar_k}}$, where $\lrange{\pvar_1}{\pvar_k}$
is the finite set of permutation variables which appear in the problem of interest.
These will be used to represent the permutative names, the name variables and the results of
applying the unknown permutations $\lrange{\pvar_1}{\pvar_k}$ to other name-terms, respectively.
For the translation we will also need additional variables to store intermediate values---for
these will use another disjoint, countably infinite set of variables $\VarTemp$.
We fix bijections into these sets, as follows:
\begin{iteMize}{$\bullet$}
  \item a bijection from $\Name$ to $\VarName$, where $\vvar{\name}\in\VarName$
    stands for $\name\in\Name$;
  \item a bijection from $\Nvar$ to $\VarNvar$, where $\vvar{\vara}\in\VarNvar$
    stands for $\vara\in\Nvar$; and
  \item for each permutation variable $\pvar$ a bijection
    from $\Name\disu\Nvar$ to $\VarPvar{\pvar}$, where
    $\pvvar{\pvar}{\vv}\in\VarPvar{\pvar}$ stands for
    $\pact{\pvar}{\vv}$ for $\vv\in\pn{\Name\disu\Nvar}$.
\end{iteMize}

\noindent For the translation we fix a trivial nominal signature
$\SIG$ where $\SIGCons{\SIG}=\SIGDtys{\SIG}=\emptyset$
and where $\SIGNtys{\SIG}=\setb{\nty}$ for some fixed name type $\nty$.
Given finite sets $\names$, $\varae$ and $\pvare$
and a finite set $\vare\subset\VarTemp$ we define a typing environment
for the corresponding non-permutative variables:
\begin{eqnarray*}
  \eunenv{\names}{\varae}{\pvare}{\vare} & \defeq &
  \setwhere{\vvar{\name}\OFTY\nty}{\name\in\names}\disu
  \setwhere{\vvar{\vara}\OFTY\nty}{\vara\in\varae}\disu \\
  &&
   \setwhere{\pvvar{\pvar}{\vv}\OFTY\nty}{\pvar\in\pvare\conj\vv\in\pn{\names\disu\varae}}\disu
  \setwhere{\var\OFTY\nty}{\var\in\vare}
\end{eqnarray*}
where $\nty$ is the single name sort from $\SIG$.
Note that $\eunenv{\names'}{\varae'}{\pvare'}{\vare'}\supseteq\eunenv{\names}{\varae}{\pvare}{\vare}$
holds if $\names'\supseteq\names$, $\varae'\supseteq\varae$, $\pvare'\supseteq\pvare$ and
$\vare'\supseteq\vare$ all hold.
The following rules specify the translation of an equivariant
unification name-term $\va$ into a variable $\var$ and associated constraints
$\prob'$ in NPNAS, which involve the new variables $\vare'$.
We write this as $\eutrnterm{\va}{\vare}{\var}{\prob'}{\vare'}$.
\begin{mathpar}
  \inferrule*{\ }
             {\eutrnterm{\pact{\idp}{\vv}}{\vare}{\vvar{\vv}}{\emptyset}{\emptyset}} \and
  \inferrule*{\ }
             {\eutrnterm{\pact{\pvar}{\vv}}{\vare}{\pvvar{\pvar}{\vv}}{\emptyset}{\emptyset}} \\
  \inferrule*{\eutrnterm{\va}{\vare}{\vvar{\va}}{\prob_{1}}{\vare_{1}} \\
              \eutrnterm{\vb}{\pn{\vare\disu\vare_{\va}}}{\vvar{\vb}}{\prob_{2}}{\vare_{2}} \\
              \vz\notin\pn{\vare\disu\vare_{1}\disu\vare_{2}}}
             {\eutrnterm{\pact{\swap{\va}{\vb}}{\vv}}{\vare}{\vz}
             {\CONJ{\prob_{1}}{\CONJ{\prob_{2}}
             {\pn{\swapcon{\vvar{\va}}{\vvar{\vb}}{\vz}{\vvar{\vv}}}}}}{\pn{\vare_{1}\disu\vare_{2}\disu\setb{\vz}}}}
\end{mathpar}
Here, and throughout, $\vare\subset\VarTemp$ is a finite set of
temporary variables which have already been used and must be
\emph{avoided} in the rest of the translation.
The following result states the semantics of the
\qtn{swapping} construction.

\begin{lem}[Swapping constraints]
  \label{lem:swap}
  Suppose that $\var,\vy,\vu,\vw\in\dom{\vln}$,
  $\vln\in\Atree{\SIG}{\etenv}$ and
  $\etenv\pn{\var}=\etenv\pn{\vy}=\nty$
  for some name sort $\nty$, and where
  $\vln\pn{\var}=\setb{\name}$ and $\vln\pn{\vy}=\setb{\name'}$,
  for some $\name$, $\name'$.
  Then we get that $\SATIS{\vln}{\swapcon{\var}{\vy}{\vu}{\vw}}$
  iff $\vln\pn{\vu}=\pact{\swap{\name}{\name'}}{\vln\pn{\vw}}$.\qed
\end{lem}

The NPNAS translation of an equivariant unification constraint
can now be defined straightforwardly.
We write $\eutrnconstr{\constr}{\vare}{\prob'}{\vare'}$ if the
equivariant unification constraint $\constr$ is translated to the
NPNAS problem $\prob'$, involving new variables $\vare'$
and avoiding $\vare$.
\begin{mathpar}
  \inferrule*{\eutrnterm{\va}{\vare}{\vvar{\va}}{\prob_{1}}{\vare_{1}} \\
              \eutrnterm{\vb}{\pn{\vare\disu\vare_{\va}}}{\vvar{\vb}}{\prob_{2}}{\vare_{2}}}
             {\eutrnconstr{\euecon{\va}{\vb}}{\vare}
             {\CONJ{\prob_{1}}{\CONJ{\prob_{2}}
             {\ECON{\vvar{\va}}{\vvar{\vb}}}}}{\pn{\vare_{1}\disu\vare_{2}}}} \\
  \inferrule*{\eutrnterm{\va}{\vare}{\vvar{\va}}{\prob_{1}}{\vare_{1}} \\
              \eutrnterm{\vb}{\pn{\vare\disu\vare_{\va}}}{\vvar{\vb}}{\prob_{2}}{\vare_{2}}}
             {\eutrnconstr{\eufcon{\va}{\vb}}{\vare}
             {\CONJ{\prob_{1}}{\CONJ{\prob_{2}}
             {\FCON{\vvar{\va}}{\vvar{\vb}}}}}{\pn{\vare_{1}\disu\vare_{2}}}}
\end{mathpar}
In order to model permutative names and permutation variables
using the standard non-permutative variables from
\Sect{\ref{sec:syntax-semantics}}, we must impose some
additional consistency constraints on the variables
to ensure that they reflect the correct semantics.
In particular, we will want to express pairwise
\emph{distinctness} constraints between finite sets of name variables.
For a finite set $\varset=\setrange{\var_1}{\var_k}$ of variables of name sort,
we write $\DISTINCT{\varset}$ for the set of atomic
freshness constraints $\setwhere{\FCON{\var_i}{\var_j}}{1\leq i<j\leq k}$.

\begin{defi}[Consistency constraints]
  \label{def:encoding-consistency}
  Given finite sets $\names$, $\varae$ and $\pvare$
  we write $\consistency{\names}{\varae}{\pvare}$
  for the \emph{consistency constraints} over
  $\names$, $\varae$ and $\pvare$, which are defined as follows.
  \begin{eqnarray*}
    \consistency{\names}{\varae}{\pvare} & \defeq &
    \CONJ{\DISTINCT{\setwhere{\vvar{\name}}{\name\in\names}}}
    {\{\bijcon{\vvar{\vv}}{\vvar{\vv'}}
    {\pvvar{\pvar}{\vv}}{\pvvar{\pvar}{\vv'}}} \\
    && \qquad\qquad\qquad\mid
    \pvar\in\pvare \conj \vv\neq\vv' \conj
    \setb{\vv,\vv'}\subseteq\pn{\names\disu\varae}\}
  \end{eqnarray*}
\end{defi}\smallskip

\noindent Consistency constraints will be crucial in the
proof that our encoding of equivariant unification is correct.
The first part of $\consistency{\names}{\varae}{\pvare}$ requires that
the variables which represent the names in $\names$ should all be
distinct---this encodes the fact that names $\name$ are permutative in
equivariant unification.
The following result states the semantics of the
$\swapcon{\vvar{\va}}{\vvar{\vb}}{\vz}{\vvar{\vv}}$
constraints used in the second part.

\begin{lem}[Bijection constraints]
  \label{lem:bij}
  Suppose that $\var,\vy,\var',\vy'\in\dom{\vln}$,
  $\vln\in\Atree{\SIG}{\etenv}$ and
  $\etenv\pn{\var}=\etenv\pn{\vy}=\etenv\pn{\var'}=\etenv\pn{\vy'}=\nty$
  for some name sort $\nty$.
  Then we get that $\SATIS{\vln}{\bijcon{\var}{\vy}{\var'}{\vy'}}$
  iff $\vln\pn{\var}=\vln\pn{\vy} \bimp \vln\pn{\var'}=\vln\pn{\vy'}$.\qed
\end{lem}

Thus the second part of $\consistency{\names}{\varae}{\pvare}$
requires that all instantiations of the variables $\pvvar{\pvar}{\vv}$
\pn{which represent the application of $\pvar$ to $\vv$}
respect the fact that $\pvar$ denotes an unknown \emph{bijection}.

We now have all the ingredients needed to define the NPNAS translation
of a name-name equivariant unification problem $\euprob$,
using similar notation to above.
\begin{mathpar}
  \inferrule*
  {\names=\namesof{\euprob} \\ \varae=\nvars{\euprob} \\ \pvare=\pvars{\euprob} \\
   \euprob\equiv\setrange{\constr_1}{\constr_k} \\\\
   \eutrnconstr{\constr_1}{\emptyset}{\prob_1}{\vare_1} \\ \cdots \\
   \eutrnconstr{\constr_k}{\pn{\disurange{\vare_1}{\vare_{k-1}}}}{\prob_k}{\vare_k}}
  {\eutrnprob{\euprob}{\CONJ{\CONJRANGE{\prob_1}{\prob_k}}
  {\consistency{\names}{\varae}{\pvare}}}{\pn{\disurange{\vare_1}{\vare_k}}}}
\end{mathpar}

\begin{lem}[Typing lemma for problem translation]
  \label{lem:eutrnprob-typing}
  If $\eujmt{\names}{\varae}{\pvare}{\euprob}$ and
  $\eutrnprob{\euprob}{\prob}{\vare}$ then
  $\formjb{\SIG}{\eunenv{\names}{\varae}{\pvare}{\vare}}{\prob}$.\qed
\end{lem}

\subsection{Correctness of the encoding}
\label{subsec:eu-correctness}

In this section we prove that every satisfying ground valuation
$\euvln$ for a name-name equivariant unification problem can be
translated into an NPNAS valuation which satisfies the corresponding
NPNAS constraint problem, and vice versa.
We begin with the translation of ground equivariant unification
\pn{EU} valuations.

\begin{defi}[Translating EU ground valuations]
  \label{def:translate-euvln}
  Given a ground valuation $\euvln$ and a finite set $\names$ of names,
  write $\euvlntrn{\euvln}{\names}$ for the NPNAS valuation where
  \begin{iteMize}{$\bullet$}
    \item $\dom{\euvlntrn{\euvln}{\names}}=\setwhere{\vvar{\name}}{\name\in\names}
      \disu\setwhere{\vvar{\vara}}{\vara\in\dom{\euvln}}
      \disu\setwhere{\pvvar{\pvar}{\vv}}{\pvar\in\dom{\euvln}\conj\vv\in\pn{\names\disu\varae}}$;
    \item $\uquant{\vvar{\name}\in\dom{\euvlntrn{\euvln}{\names}}}
      {\euvlntrn{\euvln}{\names}\pn{\vvar{\name}}=\setb{\name}}$;
    \item $\uquant{\vvar{\vara}\in\dom{\euvlntrn{\euvln}{\names}}}
      {\euvlntrn{\euvln}{\names}\pn{\vvar{\vara}}=\setb{\euvln\pn{\vara}}}$; and
    \item $\uquant{\pvvar{\pvar}{\vv}\in\dom{\euvlntrn{\euvln}{\names}}}
      {\euvlntrn{\euvln}{\names}\pn{\pvvar{\pvar}{\vv}}=
      \setb{\pact{\euvln\pn{\pvar}}{\euvln\pn{\vv}}}}$.
  \end{iteMize}\noindent
\end{defi}

\noindent The definition of $\euvlntrn{\euvln}{\names}$ uses the partitions of
$\Var$ described above to encode the different kinds of variables and names
from equivariant unification.
The additional $\names$ parameter is needed because the valuation in EU
does not provide instantiations for names, whereas the valuation in NPNAS
must provide instantiations for the variables corresponding to those names.
Note that $\euvlntrn{\euvln}{\names}\in\Atree{\SIG}{\eunenv{\names}{\varae}{\pvare}{\emptyset}}$,
where $\varae=\setwhere{\vara}{\vara\in\dom{\euvln}}$ and
$\pvare=\setwhere{\pvar}{\pvar\in\dom{\euvln}}$.
Recalling that a \qtn{vertex} $\vv$ is either a name $\name$
or a name variable $\vara$, it is straightforward to show that, for any $\vv$,
if $\namesof{\vv}\subseteq\names$ and $\varas{\vv}\subseteq\dom{\euvln}$ then
$\euvlntrn{\euvln}{\names}\pn{\vvar{\vv}}=\setb{\euvln\pn{\vv}}$.
We now show that translated EU valuations satisfy the
appropriate consistency constraints.

\begin{lem}[Translated valuations and consistency constraints]
  \label{lem:euvlntrn-consistency}
  If $\varae=\setwhere{\vara}{\vara\in\dom{\euvln}}$ and
  $\pvare=\setwhere{\pvar}{\pvar\in\dom{\euvln}}$ then
  $\SATIS{\euvlntrn{\euvln}{\names}}{\consistency{\names}{\varae}{\pvare}}$.
\end{lem}
\proof
  We must show that \textit{\pn{1}}
  if $\name\neq\name'$ and $\setb{\name,\name'}\subseteq\names$ then
  $\euvlntrn{\euvln}{\names}\pn{\vvar{\name}}\neq
   \euvlntrn{\euvln}{\names}\pn{\vvar{\name'}}$; and
  \textit{\pn{2}} if $\pvar\in\pvare$ and
  $\setb{\vv,\vv'}\subseteq\pn{\names\disu\varae}$
  then $\SATIS{\euvlntrn{\euvln}{\names}}
  {\bijcon{\vvar{\vv}}{\vvar{\vv'}}
  {\pvvar{\pvar}{\vv}}{\pvvar{\pvar}{\vv'}}}$.
  In both cases we use the definition of $\euvlntrn{\euvln}{\names}$.
  For the second we furthermore rely on the fact that $\euvln\pn{\pvar}$
  is a permutation, and hence that
  $\euvlntrn{\euvln}{\names}\pn{\vvar{\vv}}=
   \euvlntrn{\euvln}{\names}\pn{\vvar{\vv'}}$ iff
  $\euvlntrn{\euvln}{\names}\pn{\pvvar{\pvar}{\vv}}=
   \euvlntrn{\euvln}{\names}\pn{\pvvar{\pvar}{\vv'}}$,
  and the result follows by \Lem{\ref{lem:bij}}.\qed

We now show that any solution to a problem in EU can be
translated into a satisfying valuation for the
corresponding NPNAS problem.

\begin{lem}[Problem satisfaction from EU into NPNAS]
  \label{lem:prob-eu-to-npnas}
  Suppose that $\names=\namesof{\euprob}$,
  $\varae=\nvars{\euprob}=\setwhere{\vara}{\vara\in\dom{\euvln}}$ and
  $\pvare=\pvars{\euprob}=\setwhere{\pvar}{\pvar\in\dom{\euvln}}$.
  If $\SATIS{\euvln}{\euprob}$ and
  $\eutrnprob{\euprob}{\prob}{\vare}$
  then there exists a valuation
  $\twinkle{\vln}\in\Atree{\SIG}{\eunenv{\names}{\varae}{\pvare}{\vare}}$
  which extends $\euvlntrn{\euvln}{\names}$ and is such that
  $\SATIS{\twinkle{\vln}}{\prob}$.
\end{lem}
\proof
  By induction on the structure of EU name-terms $\va$, we can show that
  if $\eutrnterm{\va}{\vare}{\vz}{\prob}{\vare'}$
  \pn{where $\varae=\setwhere{\vara}{\vara\in\dom{\euvln}}$,
  $\pvare=\setwhere{\pvar}{\pvar\in\dom{\euvln}}$ and
  $\eujmt{\names}{\varae}{\pvare}{\va}$}
  then there exists a valuation
  $\twinkle{\vln}\in\Atree{\SIG}{\eunenv{\names}{\varae}{\pvare}{\pn{\vare\disu\vare'}}}$
  which extends $\euvlntrn{\euvln}{\names}$ and is such that
  $\SATIS{\twinkle{\vln}}{\prob}$
  and $\twinkle{\vln}\pn{\vz}=\setb{\euvln\pn{\va}}$.
  This relates the result of instantiating a name-term in EU
  to the corresponding term instantiation in NPNAS.

  We can then prove a similar result for atomic constraints:
  if $\SATIS{\euvln}{\constr}$ and
  $\eutrnconstr{\constr}{\vare}{\prob}{\vare'}$
  \pn{where $\varae=\setwhere{\vara}{\vara\in\dom{\euvln}}$,
  $\pvare=\setwhere{\pvar}{\pvar\in\dom{\euvln}}$ and
  $\eujmt{\names}{\varae}{\pvare}{\constr}$}
  then there exists a valuation
  $\twinkle{\vln}\in\Atree{\SIG}{\eunenv{\names}{\varae}{\pvare}{\pn{\vare\disu\vare'}}}$
  which extends $\euvlntrn{\euvln}{\names}$ and is such that
  $\SATIS{\twinkle{\vln}}{\prob}$.
  This uses the above result and relates constraint satisfaction
  in EU to constraint satisfaction in NPNAS.
  
  Then, if $\SATIS{\euvln}{\euprob}$ then $\SATIS{\euvln}{\constr}$
  holds for all $\constr\in\euprob$, where we
  suppose that $\euprob=\setrange{\constr_1}{\constr_k}$.
  We assume that $\eutrnprob{\euprob}{\prob}{\vare}$ holds,
  where $\eutrnconstr{\constr_i}{\pn{\vare\disu\disurange{\vare_1}{\vare_{i-1}}}}{\prob_i}{\vare_i}$
  holds for all $i\in\setrange{1}{k}$.
  Using the above result about constraint satisfaction
  we can construct a single NPNAS valuation
  $\twinkle{\vln}$ which extends $\euvlntrn{\euvln}{\names}$
  and is such that $\SATIS{\twinkle{\vln}}{\prob_i}$ holds for all $i\in\setrange{1}{k}$.
  Thus we have that $\SATIS{\twinkle{\vln}}{\CONJRANGE{\prob_1}{\prob_k}}$ holds.
  By \Lem{\ref{lem:euvlntrn-consistency}} we know that
  $\SATIS{\euvlntrn{\euvln}{\names}}{\consistency{\names}{\varae}{\pvare}}$
  holds: hence $\SATIS{\twinkle{\vln}}{\consistency{\names}{\varae}{\pvare}}$ holds also.
  Hence we get that $\SATIS{\twinkle{\vln}}{\prob}$ holds, as required.\qed

We now turn to the other direction---we begin by proving that
any NPNAS valuation which satisfies a set of consistency constraints
can be translated back into a corresponding EU valuation.

\begin{lem}[Consistency constraints imply an EU valuation]
  \label{lemma:consistency-euvln}
  If $\vln\in\Atree{\SIG}{\eunenv{\names}{\varae}{\pvare}{\vare}}$
  and $\SATIS{\vln}{\consistency{\names}{\varae}{\pvare}}$ then
  there exists a permutation $\perm_{\vln}$ and a ground EU valuation
  $\euvln_{\vln}$ \pn{with $\pn{\varae\disu\pvare}\subseteq\dom{\euvln_{\vln}}$} such that
  \begin{enumerate}[\em(1)]
    \item $\uquant{\name\in\names}{\vln\pn{\vvar{\name}}=\setb{\perm_{\vln}\pn{\name}}}$;
    \item $\uquant{\vara\in\varae}{\vln\pn{\vvar{\vara}}=\setb{\pact{\perm_{\vln}}{\pn{\euvln_{\vln}\pn{\vara}}}}}$; and
    \item $\uquant{\pvar\in\pvare}{\uquant{\vv\in\pn{\names\disu\varae}}{\vln\pn{\pvvar{\pvar}{\vv}}=
          \setb{\pact{\perm_{\vln}}{\pn{\pn{\euvln_{\vln}\pn{\pvar}}\pn{\euvln_{\vln}\pn{\vv}}}}}}}$.
  \end{enumerate}
\end{lem}
\proof
  We prove the three points separately.
  \begin{enumerate}[(1)]

    \item This follows from the fact that
      $\SATIS{\vln}{\DISTINCT{\setwhere{\vvar{\name}}{\name\in\names}}}$ holds,
      which implies that we can fix a permutation $\perm_{\vln}$ which is a bijection
      between $\names$ and $\setwhere{\vln\pn{\vvar{\name}}}{\name\in\names}$, as required.

    \item We note that $\setwhere{\vvar{\vara}}
      {\vara\in\varae}\inter\setwhere{\vvar{\name}}{\name\in\names}=\emptyset$.
      Thus we can construct a ground EU valuation $\euvln_{\vln}$ such that
      $\uquant{\vara\in\varae}{\vln\pn{\vvar{\vara}}=
      \setb{\pact{\perm_{\vln}}{\pn{\euvln_{\vln}\pn{\vara}}}}}$,
      by setting $\euvln_{\vln}\pn{\vara}=\pinv{\perm_{\vln}}\pn{\name}$
      if $\vln\pn{\vvar{\vara}}=\setb{\name}$.

    \item Since $\SATIS{\vln}{\consistency{\names}{\varae}{\pvare}}$ we know that
      $\SATIS{\vln}{\bijcon{\vvar{\vv}}{\vvar{\vv'}}{\pvvar{\pvar}{\vv}}{\pvvar{\pvar}{\vv'}}}$
      holds for all $\pvar\in\pvare$ and all $\vv,\vv'\in\pn{\names\disu\varae}$.
      By \Lem{\ref{lem:bij}} we get that there is a bijection between
      $\setwhere{\vln\pn{\vvar{\vv}}}{\vv\in\pn{\names\disu\varae}}$ and
      $\setwhere{\vln\pn{\pvvar{\pvar}{\vvar{\vv}}}}{\vv\in\pn{\names\disu\varae}}$.
      We represent this bijection as a permutation $\perm$.
      From the first two proof obligations we know that
      $\vln\pn{\vvar{\vv}}=\setb{\pact{\perm_{\vln}}{\pn{\euvln_{\vln}\pn{\vv}}}}$
      for all $\vv\in\pn{\names\disu\varae}$, and hence that
      $\vln\pn{\pvvar{\pvar}{\vv}}=\setb{\pact{\perm}{\pn{\pact{\perm_{\vln}}{\pn{\euvln_{\vln}\pn{\vv}}}}}}$, i.e. that
      $\vln\pn{\pvvar{\pvar}{\vv}}=\setb{\pact{\pn{\pcomp{\perm}{\perm_{\vln}}}}{\pn{\euvln_{\vln}\pn{\vv}}}}$,
      for all $\vv\in\pn{\names\disu\varae}$.
      Now, we can decompose $\perm$ into the form $\pcomps{\perm_{\vln}}{\perm_{\pvar}}{\pinv{\perm_{\vln}}}$
      for some $\perm_{\pvar}$.
      Thus we get that, for all $\vv\in\pn{\names\disu\varae}$, $\vln\pn{\pvvar{\pvar}{\vv}}=
      \setb{\pact{\pn{\pcompss{\perm_{\vln}}{\perm_{\pvar}}{\pinv{\perm_{\vln}}}{\perm_{\vln}}}}{\pn{\euvln_{\vln}{\vv}}}}=
      \setb{\pact{\pn{\pcomp{\perm_{\vln}}{\perm_{\pvar}}}}{\pn{\euvln_{\vln}{\vv}}}}=
      \setb{\pact{\perm_{\vln}}{\pn{\pact{\perm_{\pvar}}{\pn{\euvln_{\vln}\pn{\vv}}}}}}$.
      Thus if we let $\euvln_{\vln}\pn{\pvar}=\perm_{\pvar}$ then we get that
      $\uquant{\pvar\in\pvare}{\uquant{\vv\in\pn{\names\disu\varae}}{\vln\pn{\pvvar{\pvar}{\vv}}=
      \setb{\pact{\perm_{\vln}}{\pn{\pn{\euvln_{\vln}\pn{\pvar}}\pn{\euvln_{\vln}\pn{\vv}}}}}}}$ holds, as required.\qed\smallskip

  \end{enumerate}

\noindent We now show that any satisfying valuation for the
NPNAS translation of an EU problem can be translated back
into a solution of the original EU problem.

\begin{lem}[Problem satisfaction from NPNAS into EU]
  \label{lemma:prob-npnas-to-eu}
  Suppose that $\names=\namesof{\euprob}$, $\varae=\nvars{\euprob}$ and
  $\pvare=\pvars{\euprob}$, and that $\eutrnprob{\euprob}{\prob}{\vare}$,
  $\vln\in\Atree{\SIG}{\eunenv{\names}{\varae}{\pvare}{\vare}}$ and
  $\SATIS{\vln}{\prob}$ all hold.
  Then there exists a ground EU valuation $\euvln_{\vln}$
  \pn{with $\pn{\varae\disu\pvare}\subseteq\dom{\euvln_{\vln}}$}
  such that $\SATIS{\euvln_{\vln}}{\euprob}$.
\end{lem}
\proof
  The structure of this proof rather mirrors that of
  \Lem{\ref{lem:prob-eu-to-npnas}}.
  We begin by showing that if $\eujmt{\names}{\varae}{\pvare}{\va}$,
  $\eutrnterm{\va}{\vare}{\vz}{\prob}{\vare'}$,
  $\vln\in\Atree{\SIG}{\eunenv{\names}{\varae}{\pvare}{\pn{\vare\disu\vare'}}}$
  and $\SATIS{\vln}{\prob}$ all hold then
  $\vln\pn{\vz}=\setb{\pact{\perm_{\vln}}{\euvln_{\vln}\pn{\va}}}$ holds,
  for some $\euvln_{\vln}$ and $\perm_{\vln}$ which satisfy the three conditions
  stated in \Lem{\ref{lemma:consistency-euvln}}.
  This relates instantiations of translated EU name-terms in
  NPNAS back to instantiations of the original EU term,
  up to a permutation.

  We proceed to prove a similar result about constraint
  satisfaction---if $\eujmt{\names}{\varae}{\pvare}{\constr}$,
  $\eutrnconstr{\constr}{\vare}{\prob}{\vare'}$,
  $\vln\in\Atree{\SIG}{\eunenv{\names}{\varae}{\pvare}{\pn{\vare\disu\vare'}}}$ and
  $\SATIS{\vln}{\prob}$ all hold then $\SATIS{\euvln_{\vln}}{\constr}$,
  where $\euvln_{\vln}$ satisfies the three conditions
  stated in \Lem{\ref{lemma:consistency-euvln}} \pn{for some $\perm_{\vln}$}.
  This uses the previous result and shows that satisfaction of
  translated EU constraints in NPNAS can be related back to the
  semantics of the original constraint in EU.
  There is a twist here, as we must use the equivariance property
  of the NPNAS semantics \pn{\Rem{\ref{rem:permutations}}}
  to strip off the unwanted permutation $\perm_{\vln}$.
  
  Thus, if $\eutrnprob{\euprob}{\prob}{\vare}$ then
  $\prob\equiv\pn{\CONJ{\CONJRANGE{\prob_1}{\prob_k}}
  {\consistency{\names}{\varae}{\pvare}}}$, $\euprob\equiv\setrange{\constr_1}{\constr_k}$
  and $\vare=\disurange{\vare_1}{\vare_k}$ all hold, where
  $\names=\namesof{\euprob}$, $\varae=\nvars{\euprob}$, $\pvare=\pvars{\euprob}$ and where
  $\eutrnconstr{\constr_i}{\pn{\disurange{\vare_1}{\vare_{i-1}}}}{\prob_i}{\vare_i}$
  holds for all $i\in\setrange{1}{k}$.
  It follows that $\eujmt{\names}{\varae}{\pvare}{\euprob}$ holds.
  We assume that $\SATIS{\vln}{\prob}$, i.e.~that
  $\SATIS{\vln}{\consistency{\names}{\varae}{\pvare}}$ holds and that
  $\SATIS{\vln}{\prob_i}$ holds for all $i\in\setrange{1}{k}$.
  By \Lem{\ref{lemma:consistency-euvln}} we get that there exists
  a permutation $\perm_{\vln}$ and a ground EU valuation $\euvln_{\vln}$
  \pn{with $\pn{\varae\disu\pvare}\subseteq\dom{\euvln_{\vln}}$} which
  satisfy the three conditions laid out in the statement of that lemma.
  Then, using the above result on constraint satisfaction, we can show that
  $\SATIS{\euvln_{\vln}}{\constr_i}$ holds for all $i\in\setrange{1}{k}$,
  i.e.~that $\SATIS{\euvln_{\vln}}{\euprob}$ holds, as required.\qed

The key result of this section is the following theorem,
which demonstrates the correctness of the encoding of
name-name equivariant unification into non-permutative
nominal constraints.

\begin{thm}[Correctness of encoding]
  \label{thm:corr}
  Suppose that $\eujmt{\names}{\varae}{\pvare}{\euprob}$
  and $\eutrnprob{\euprob}{\prob}{\vare}$ both hold.
  Then, $\euprob$ is satisfiable iff
  $\cprob{\eunenv{\names}{\varae}{\pvare}{\vare}}{\prob}$
  is satisfiable.
\end{thm}
\proof
  For the forward direction: 
  if $\SAT{\euprob}$ then there exists a ground EU valuation
  $\euvln$ such that $\SATIS{\euvln}{\euprob}$.
  If $\eutrnprob{\euprob}{\prob}{\vare}$ then
  by \Lem{\ref{lem:prob-eu-to-npnas}} there exists a valuation
  $\twinkle{\vln}\in\Atree{\SIG}{\eunenv{\names}{\varae}{\pvare}{\vare}}$
  such that $\SATIS{\twinkle{\vln}}{\euprob}$,
  from which it follows that $\SAT{\euprob}$ holds.
  For the reverse direction:
  if $\SAT{\prob}$ then there exists a valuation
  $\vln\in\Atree{\SIG}{\eunenv{\names}{\varae}{\pvare}{\vare}}$
  such that $\SATIS{\vln}{\prob}$.
  By \Lem{\ref{lemma:prob-npnas-to-eu}} we can construct
  a ground EU valuation $\euvln_{\vln}$ such that
  $\SATIS{\euvln_{\vln}}{\euprob}$, which shows that
  $\SAT{\euprob}$ holds.\qed

\begin{exa}[A translated EU problem]
\label{ex:eu-problem}
  As an example, consider the following equivariant unification problem
  which involves permutation variables and swappings.
  \begin{equation*}
    \euprob=\setb{\ECON{\pn{\pact{\pvar}{\vara}}}{\pact{\swap{\pn{\pact{\pvar}{\vara}}}{\pn{\pact{\pvar}{\varb}}}}{\pn{\pact{\pvar}{\vara}}}},
    \FCON{\pn{\pact{\pvar'}{\vara}}}{\pn{\pact{\pvar'}{\varb}}}}
  \end{equation*}
  We would expect this to be unsatisfiable because the first constraint implies that
  $\vara=\varb$ whereas the second implies that $\vara\neq\varb$.
  The translation of $\euprob$ into NPNAS is as follows, where $\vz$ is a freshly chosen variable.
  \begin{eqnarray}
    \label{eq:ex1}
    && \swapcon{\pvvar{\pvar}{\vara}}{\pvvar{\pvar}{\varb}}{\vz}{\pvvar{\pvar}{\vara}} \\
    \label{eq:ex2}
    & \AND & \ECON{\pvvar{\pvar}{\vara}}{\vz} \\
    \label{eq:ex3}
    & \AND & \FCON{\pvvar{\pvar'}{\vara}}{\pvvar{\pvar'}{\varb}} \\
    \label{eq:ex4}
    & \AND & \bijcon{\vvar{\vara}}{\vvar{\varb}}{\pvvar{\pvar}{\vara}}{\pvvar{\pvar}{\varb}} \\
    \label{eq:ex5}
    & \AND & \bijcon{\vvar{\vara}}{\vvar{\varb}}{\pvvar{\pvar'}{\vara}}{\pvvar{\pvar'}{\varb}}
  \end{eqnarray}
  To see that the NPNAS problem is also unsatisfiable, we observe that
  \ref{eq:ex1} and \ref{eq:ex2} imply that $\ECON{\pvvar{\pvar}{\vara}}{\pvvar{\pvar}{\varb}}$.
  This fact, along with \ref{eq:ex4}, implies that $\ECON{\vvar{\vara}}{\vvar{\varb}}$ which,
  together with \ref{eq:ex5}, implies that $\ECON{\pvvar{\pvar'}{\vara}}{\pvvar{\pvar'}{\varb}}$ holds.
  However, this contradicts \ref{eq:ex3} and thus it follows that the NPNAS problem is unsatisfiable.

  In solving this problem, the decision procedure outlined in the proof of \Thm{\ref{thm:alg-decid}}
  must first construct and solve the first-order reduction of the NPNAS problem,
  as a termination check.
  In this case, this is straightforward as the names are erased and the abstractions
  replaced by tuples as outline above.
  Writing $\UNITS{k}{\term}$ for
  $\PAIR{\UNIT}{\PAIR{\UNIT}{\cdots\PAIR{\UNIT}{\term}}}$
  if there are $k$ nested occurrences of $\UNIT$, this leaves
  the following first-order unification problem which is trivially satisfiable.
  \begin{equation*}
    \pn{\ECON{\UNITS{2}{\UNIT}}{\UNITS{2}{\UNIT}}} \AND
    \pn{\ECON{\UNIT}{\UNIT}} \AND
    \pn{\ECON{\UNITS{2}{\UNIT}}{\UNITS{2}{\UNIT}}} \AND
    \pn{\ECON{\UNITS{2}{\UNIT}}{\UNITS{2}{\UNIT}}}
  \end{equation*}
  Having ascertained that the problem is strongly normalising, we proceed to compute the set of
  $\trans$-normal forms using the reduction rules from \Fig{\ref{fig:trans}}.
  There are 27 cases to check in total, since for each of \ref{eq:ex1}, \ref{eq:ex4} and \ref{eq:ex5}
  there are 3 branches according to reduction rule \erule{4}.
  We do not specify a particular search strategy, provided that the entire reduction
  space is explored.
  In this case, none of the $\trans$-normal forms turn out to be solved in the sense of
  \Defn{\ref{def:constraint-solved-prob}}, which corresponds to the fact that the problem is unsatisfiable,
  as argued above.
\end{exa}

To see that the encoding presented in this section is a
polynomial time reduction, suppose that there are $\numof{\name}$ names,
$\numof{\vara}$ name variables, $\numof{\pvar}$ permutation variables,
$\numof{\mathit{swap}}$ swappings and $\numof{\constr}$ constraints in $\euprob$.
Then there are $\numof{\name}\pn{\numof{\name}-1} +
\numof{\pvar}\pn{\numof{\name}+\numof{\vara}} +
\numof{\constr} + \numof{\mathit{swap}}$ constraints and
$\numof{\name} + \numof{\vara} + \numof{\pvar}\pn{\numof{\name} +
\numof{\vara}} + \numof{\mathit{swap}}$ variables in $\prob$,
where $\eutrnprob{\euprob}{\prob}{\vare}$ \pn{for some $\vare$}.
These are both polynomial functions of the size of $\euprob$.

Note that the translation defined in this section only deals with
name-name equivariant unification problems.
We can use the results from this section to derive a translation of
full equivariant unification into NPNAS by using the \qtn{first phase}
of Cheney's algorithm \cite[\Sect{4.2.1}]{CheneyJ:equu-jv} to reduce the
problem into a name-name problem \pn{if possible} before using the algorithm
described herein to translate it into NPNAS.
Note, however, that this is not a polynomial time reduction because
the \qtn{first phase} of Cheney's algorithm has an exponential upper bound.

\begin{rem}[A tractable subproblem]
\label{rem:tractable}
  Say that a non-permutative constraint problem
  $\cprob{\etenv}{\prob}$ is \emph{permutative} iff every valuation
  $\vln\in\Atree{\SIG}{\etenv}$ such that
  $\SATIS{\vln}{\prob}$ also has
  $\SATIS{\vln}{\DISTINCT{\nvars{\etenv}}}$,
  where $\nvars{\etenv}$ is the set of all variables
  $\var$ in $\dom{\etenv}$ such that $\etenv\pn{\var}$
  is a name sort.
  It was shown in \cite[\Sect{6.4}]{LakinMR:exemli} that
  satisfiability of problems in this subset can be decided
  in polynomial time via translation to nominal unification.
  This makes sense because the non-permutative behaviour
  which provides the additional power in NPNAS has been disallowed.
\end{rem}

\section{Related and future work}
\label{sec:related-future}

We have already discussed at length the relationship
between non-permutative nominal constraints and the
equivariant unification problem.
The nominal unification problem of Urban, Pitts and Gabbay
\cite{PittsAM:nomu-jv} can be thought of as a ground
subproblem of equivariant unification and hence
its relationship to the work reported in this paper
is subsumed by that of equivariant unification.

Our algorithm bears some similarities to Huet's algorithm for
higher-order unification, that is, unification for
typed $\lambda$-terms \cite{HuetG:uniatl}.
That algorithm ignores equations between two terms
on the basis that they are always satisfiable,
much as our constraint transformation procedure
ignores constraints of the form
$\ECON{\abss{\var}{1}{k}{\var}}{\abss{\vy}{1}{k}{\vy}}$
\pn{where $\var$ and $\vy$ are not of name sort}.
There may be other parallels worthy of investigation.
However, it is worth noting that higher-order unification
is known to be undecidable \cite{GoldfarbWD:undsou}
whereas we have shown that a decision procedure
exists for satisfiability of non-permutative nominal
constraints \pn{see \Thm{\ref{thm:alg-decid}}}.

Higher-order unification forms the basis of an alternative technique for
representing abstract syntax with binders known as
higher-order abstract syntax \pn{HOAS} \cite{PfenningF:hoas}
which has been used in various tools for specifying,
and reasoning about, formal systems with binding
constructs \cite{NadathurG:ovelp,PfenningF:twemlf}.
These tools often exploit higher-order patterns,
which are a restricted class of $\lambda$-terms for
which unification \pn{modulo $\alpha\beta_0\eta$-equivalence}
is decidable \cite{MillerD:logpll}.
Higher-order pattern unification has been
shown to be equivalent to nominal unification
\cite{CheneyJ:relnho,LevyJ:nomuho} and it follows that
our non-permutative nominal constraint language subsumes
higher-order pattern unification just as it does
nominal unification.

Future work is needed on the relationship between non-permutative
nominal constraints and the full equivariant unification problem
\pn{i.e.~not just for name terms}.
This may involve finding an equivalent
of the bijection constraint construction from \Lem{\ref{lem:bij}}
which works for variables of any type, not just name sorts.
The termination checker described above could be
run initially or in parallel with the constraint transformation
algorithm, or omitted altogether to give a semi-decision procedure.
An alternative implementation strategy could be to encode
non-permutative constraint problems as boolean formulae and
use a SAT solver to decide their satisfiability.
With more work it may be possible to improve the algorithm for
deciding satisfiability of constraint problems in NPNAS,
in particular with regard to termination and
non-deterministic search.
The ideal algorithm would avoid the use of name-swappings
and not require a separate termination check.

A key motivation for the development of equivariant
unification was to give complete implementations of
nominal logic programming \cite{CheneyJ:nomlp-jv}
and nominal rewriting \cite{FernandezM:nomr-jv}
in cases where the nominal unification
\cite{PittsAM:nomu-jv} is not sufficiently powerful.
In other work \cite{LakinMR:residb,LakinMR:exemli}
we have investigated the use of non-permutative
nominal constraints in the context of the
functional-logic programming language \aML{}---further
work may be to investigate the theory of
rewriting over non-permutative nominal terms.

\section{Conclusion}
\label{sec:conc}

Non-permutative nominal abstract syntax is a means of encoding
terms with binders without the need for globally-fresh
permutative names.
We have defined a semantics for equality and freshness
constraints over non-permutative nominal terms,
and presented an algorithm for deciding satisfiability
of these constraint problems.
Our constraint solving procedure is novel in that it
does not use permutations, which are standard in most
nominal approaches to abstract syntax.
This simplifies the term language but complicates the analysis,
in particular the proof of termination.
Our translation of name-name equivariant unification problems
into non-permutative nominal constraints is also novel
and demonstrates explicitly how the additional features of
equivariant unification can be encoded using just
permutative variables in binding position.
Studies of non-permutative nominal constraints are
important from both a theoretical and a practical perspective,
as this algorithm could be used instead of the
more complicated equivariant unification algorithm in
situations where nominal unification cannot compute
all solutions.

\section*{Acknowledgements}
The author would like to thank Andrew Pitts for
invaluable discussions on the subject matter of this paper,
as well as Paul Blain Levy and an anonymous reviewer
for help debugging proofs.

\bibliographystyle{alpha}
\bibliography{refs}

\end{document}